\documentclass[fleqn,10pt]{wlscirep}
\usepackage[T1]{fontenc}
\usepackage{geometry}
\usepackage[utf8]{inputenc}
\usepackage{amsmath}
\usepackage{amssymb}
\usepackage{amsthm}
\usepackage{hyperref}
\usepackage{cleveref}
\usepackage{empheq}
\usepackage{enumitem}
\setlist{nosep}
\usepackage{comment}
\usepackage{hyperref}
\usepackage{subcaption} 
\usepackage{soul}

\usepackage{tikz}
\usetikzlibrary{calc}
\usetikzlibrary{shapes.geometric}
\usetikzlibrary{calc}
\newlength{\tfwidth}
\newlength{\tfheight}
\newlength{\tfxa}
\newlength{\tfxb}
\newlength{\tfya}
\newlength{\tfyb}
\newcommand{\trimFigWithBox}[6]{%
\setlength\fboxsep{0pt}%
\setlength\fboxrule{1.0pt}
\fbox{\includegraphics[width=#2, clip, trim=#3 #4 #5 #6]{#1}}%
}
\newcommand{\trimFigNoBox}[6]{%
\setlength\fboxsep{1pt}
\setlength\fboxrule{0.0pt}
\fbox{\includegraphics[width=#2, clip, trim=#3 #4 #5 #6]{#1}}%
}
\newcommand{\trimFigHeightWithBox}[6]{%
\setlength\fboxsep{0pt}%
\setlength\fboxrule{1.0pt}
\fbox{\includegraphics[height=#2, clip, trim=#3 #4 #5 #6]{#1}}%
}
\newcommand{\trimFigHeightNoBox}[6]{%
\setlength\fboxsep{1pt}
\setlength\fboxrule{0.0pt}
\fbox{\includegraphics[height=#2, clip, trim=#3 #4 #5 #6]{#1}}%
}

\newsavebox\figBox

\newcommand{\trimw}[6]{%
\sbox\figBox{\includegraphics{#1}}
\setlength{\tfwidth}{\the\wd\figBox}
\setlength{\tfheight}{\the\ht\figBox}
\setlength{\tfxa}{\tfwidth*\real{#3}}%
\setlength{\tfxb}{\tfwidth*\real{#4}}%
\setlength{\tfya}{\tfheight*\real{#5}}%
\setlength{\tfyb}{\tfheight*\real{#6}}%
\trimFigNoBox{#1}{#2}{\tfxa}{\tfya}{\tfxb}{\tfyb}%
}

\newcommand{\trimwb}[6]{%

\sbox\figBox{\includegraphics{#1}}
\setlength{\tfwidth}{\the\wd\figBox}
\setlength{\tfheight}{\the\ht\figBox}
\setlength{\tfxa}{\tfwidth*\real{#3}}%
\setlength{\tfxb}{\tfwidth*\real{#4}}%
\setlength{\tfya}{\tfheight*\real{#5}}%
\setlength{\tfyb}{\tfheight*\real{#6}}%
\trimFigWithBox{#1}{#2}{\tfxa}{\tfya}{\tfxb}{\tfyb}%
}

\newcommand{\trimh}[6]{%
\sbox\figBox{\includegraphics{#1}}
\setlength{\tfwidth}{\the\wd\figBox}
\setlength{\tfheight}{\the\ht\figBox}
\setlength{\tfxa}{\tfwidth*\real{#3}}%
\setlength{\tfxb}{\tfwidth*\real{#4}}%
\setlength{\tfya}{\tfheight*\real{#5}}%
\setlength{\tfyb}{\tfheight*\real{#6}}%
\trimFigHeightNoBox{#1}{#2}{\tfxa}{\tfya}{\tfxb}{\tfyb}%
}

\newcommand{\trimhb}[6]{%

\sbox\figBox{\includegraphics{#1}}
\setlength{\tfwidth}{\the\wd\figBox}
\setlength{\tfheight}{\the\ht\figBox}
\setlength{\tfxa}{\tfwidth*\real{#3}}%
\setlength{\tfxb}{\tfwidth*\real{#4}}%
\setlength{\tfya}{\tfheight*\real{#5}}%
\setlength{\tfyb}{\tfheight*\real{#6}}%
\trimFigHeightWithBox{#1}{#2}{\tfxa}{\tfya}{\tfxb}{\tfyb}%
}
\usepackage{xcolor, soul}

\newcommand{\bogus}[1]{}

\title{Physics consistent machine learning framework for inverse modeling with applications to ICF capsule implosions}

\author[1,*]{Daniel A. Serino}
\author[1]{Evan Bell}%
\author[1]{Marc Klasky}%
\author[1]{Ben S. Southworth}%
\author[2]{Balasubramanya Nadiga}
\author[3]{Trevor Wilcox}
\author[1]{Oleg Korobkin}
\affil[1]{Theoretical Division, Los Alamos National Laboratory, P.O. Box 1663, Los Alamos, NM 87545 U.S.}
\affil[2]{Computer, Computational, and Statistical Sciences Division, Los Alamos National Laboratory, P.O. Box 1663, Los Alamos, NM 87545 U.S.}
\affil[3]{Theoretical Design Division, Los Alamos National Laboratory, P.O. Box 1663, Los Alamos, NM 87545 U.S.}
\affil[*]{dserino@lanl.gov}

\date{\today}

\begin{abstract}
In high energy density physics (HEDP) and inertial confinement fusion (ICF), predictive modeling is complicated by uncertainty in parameters that characterize various aspects of the modeled system, such as those characterizing material properties, equation of state (EOS), opacities, and initial conditions. Typically, however, these parameters are not directly observable. What is observed instead is a time sequence of radiographic projections using X-rays. In this work, we define a set of sparse hydrodynamic features derived from the outgoing shock profile and outer material edge, which can be obtained from radiographic measurements, to directly infer such parameters. 
Our machine learning (ML)-based methodology involves a pipeline of 
two architectures, a radiograph-to-features network (R2FNet) and
a features-to-parameters network (F2PNet), that are trained independently 
and later combined to approximate a posterior distribution for the parameters
from radiographs.
We show that the machine learning architectures are able to accurately infer initial conditions and EOS parameters, and that the estimated parameters can be used in a hydrodynamics code to obtain density fields, shocks, and material interfaces that satisfy \emph{thermodynamic and hydrodynamic consistency}. 
Finally, we demonstrate that features resulting from an unknown EOS model can be successfully mapped onto parameters of a chosen analytical EOS model, implying that network predictions are learning physics, with a degree of invariance to the underlying choice of EOS model.  
To the best of our knowledge, our framework is the first demonstration of recovering both
thermodynamic and hydrodynamic consistent density fields from noisy radiographs.
\end{abstract}

\begin{document}


\maketitle


\section{Introduction}

\subsection{Inferring physical parameters from radiographic data}

Simulation plays a major role in the experimental design and analysis of radiation hydrodynamic behavior in high energy density physics (HEDP) and inertial confinement fusion (ICF), e.g. \cite{gittings2008rage,fryxell2000flash,marinak1998comparison,zimmerman1977lasnex,keller1999draco,haines2022development}, as well as in the discovery of material properties of objects made to undergo strong deformations in material science and shock physics \cite{mcglaun1990cth,weseloh2011pagosa,summers1997alegra}.  Improving the predictive skill of these simulations is likely to be key in ensuring continued progress in the design of robust burning ICF capsules \cite{abu2022lawson,weber2017improving}. However, in many of these applications predictive modeling is complicated by the inherent uncertainty in parameters used to model material properties, equation of state (EOS), opacities, and constitutive relationships, as well as complex physics associated with the initial conditions, e.g., the laser drive in ICF experiments \cite{gaffney2018review,lindl1993icf,rosen1995physics,lee2019effects,rosen1999physics,thomas2012drive}. Furthermore, understanding uncertainties in the initial conditions as well as the role of measurement uncertainty are both crucial in improving the predictive behavior of simulations. Consequently, dynamic experimentation plays a crucial role in calibrating models to improve simulations of hydrodynamic behavior and the discovery of material properties.

Historically to investigate material properties such as EOS and constitutive relationships, e.g., material strength in shock physics and material science, an impulse-response approach is used wherein the velocity trace response of the material specimen to an impulse is measured using velocity interferometry \cite{barker1972laser,malone2006overview,celliers2004line}. Indeed, the development of laser interferometry enabled the time-resolved measurement of the velocity from a reflecting surface \cite{barker1972laser}. This allowed for the measurement of the free surface and window interface velocities in dynamic compression experiments \cite{mccoy2017lagrangian}. These measurements have yielded valuable data on compressive behavior and strength of materials during both shock compression and release \cite{mccoy2017lagrangian,ahrens1968material,asay1978self,lipkin1977reshock,brown2013extracting}. Characterization of  Rayleigh-Taylor (RT) 
and Richtmyer-Meshkov (RM) 
instabilities, e.g., in terms of perturbation growth rates, are also widely used to examine constitutive relationships in shock physics, and  to quantify material strength in materials undergoing extreme deformation \cite{barnes1974taylor,colvin2003model,barton2011multiscale,smith2011high,piriz2008richtmyer,piriz2009richtmyer,dimonte2011use,buttler2012unstable,ortega2014numerical,mikaelian2013shock,plohr2005linearized,prime2016using}. Furthermore, these investigations enable examinations of asymmetry in geometric perturbations due to, e.g., manufacturing, as well as velocity perturbations, i.e drive asymmetries, to both understand and design control strategies to 
minimize the resulting hydrodynamic instabilities that degrade ICF performance \cite{thomas2012drive,thomas2013drive,weber2020mixing,delorme2015experimental,zhou2019turbulent,proano2017toward}.

\begin{figure*}[tbh]
\resizebox{\textwidth}{!}{%
  \begin{tikzpicture}
  \useasboundingbox (0,-0.5) rectangle (15,4.25);  

  \draw(-.5, .5) node[anchor=south west] {\trimw{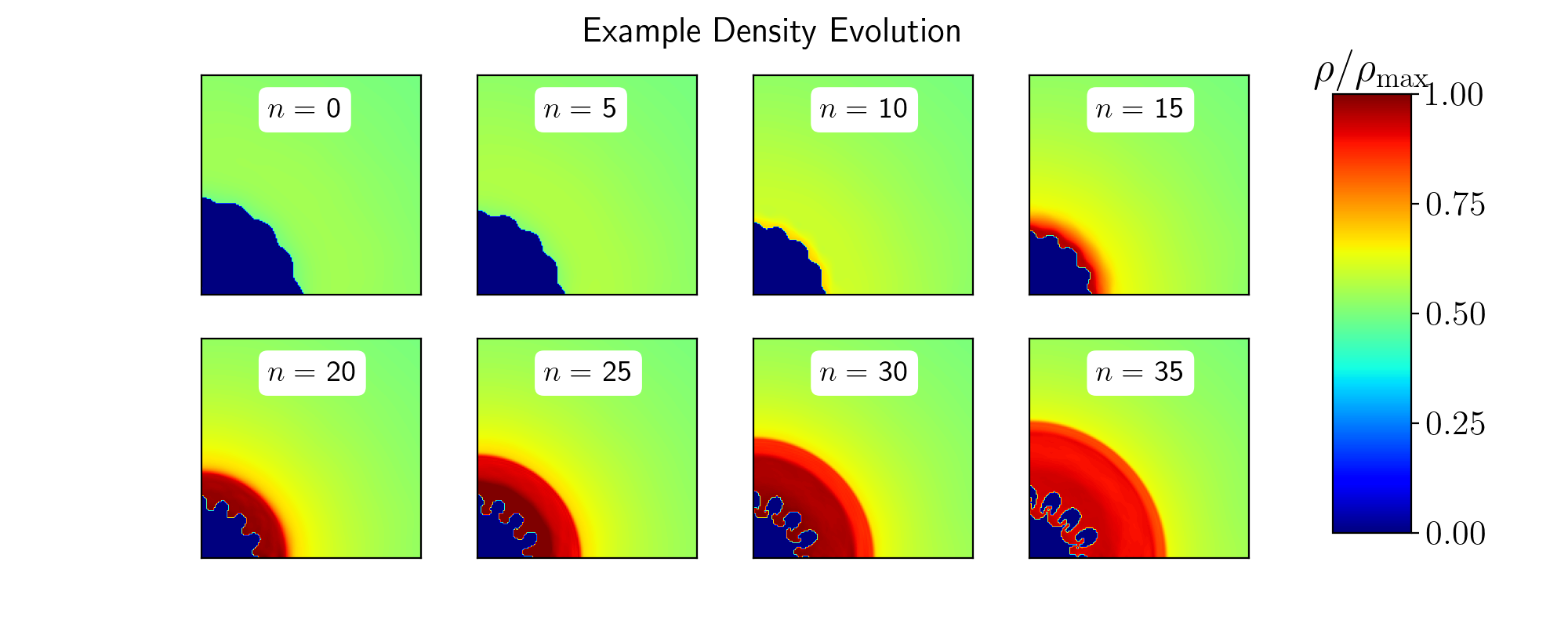}{7.4cm}{.1}{.025}{.1}{0}};
  \draw(3.25, 4) node[] {(a)};

  \draw(15, -.5) node[anchor=south east] {\trimw{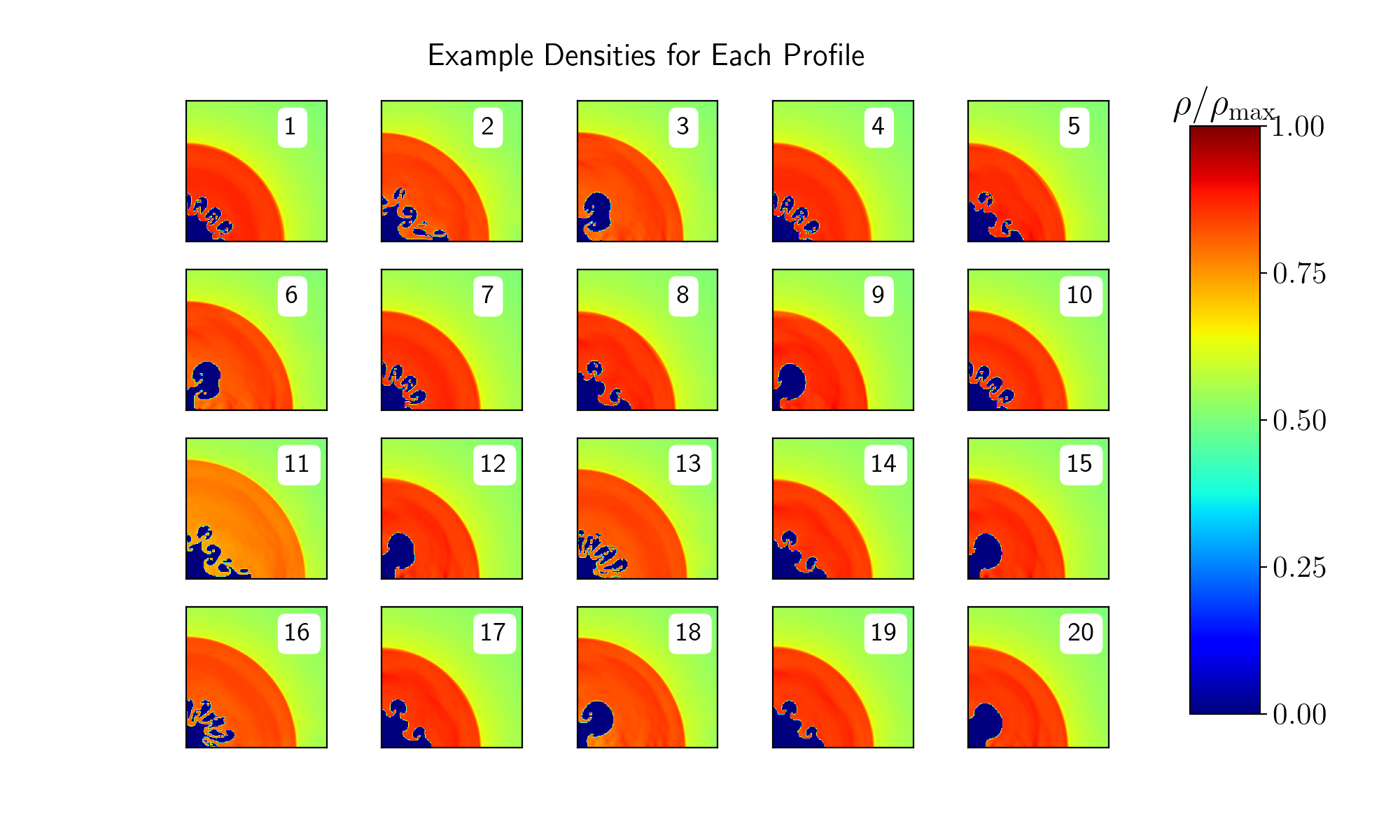}{7.4cm}{.1}{.025}{.1}{0}};
    \draw(10.75, 4.25) node[] {(b)};
  \end{tikzpicture}
}
    \caption{Example plots of the density evolution (a) and the various RMI profiles representing each inner surface perturbation profile at a fixed time frame $n=40$ (b). The images are 150x150 pixels representing the domain $\left[0, \frac{15}{44} L\right] \times \left[0, \frac{15}{44} L\right]$.}
    \label{fig:dyn_egs}
\end{figure*}

\begin{figure*}[!htb]
  \centering
  \includegraphics[trim=4cm 3cm 2.7cm 2cm, clip,width=.9\textwidth]{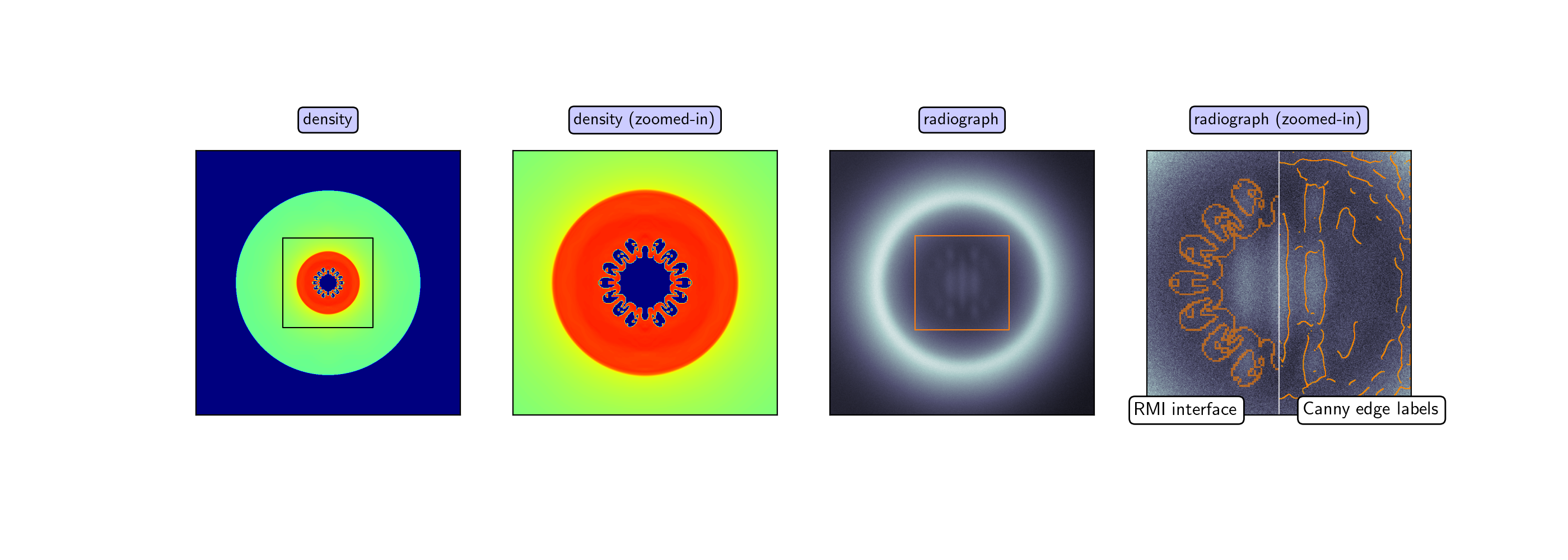}
  \caption{Sample $(r, z)$ projection of the density (1$^{\rm st}$ column), 
  zoomed-in view of the Richtmyer-Meshkov interface (2$^{\rm nd}$ column),
  synthetic radiograph (3$^{\rm rd}$ columns), 
  and a zoomed-in view of the radiograph (4$^{\rm th}$ columns)
  labeled with the RMI interface (left half) 
  and Canny edge labels (right half).
  }
  \label{fig:sampledensitysingle}
\end{figure*}

To examine RT and RM instabilities in extreme temperature and pressure conditions in a laboratory setting, spherical convergent geometries are necessary. These extreme conditions play a fundamental role in the design of robust burning ICF capsules.  Figure~\ref{fig:dyn_egs}
presents a time-history of an evolving instability in which an outward going shock interacts with a perturbed surface which gives rise to a RM instability. In this setting radiographic measurements serve as the primary means to identify key features such as peaks and troughs and to quantify empirical growth rates, which then serve as reference data to validate theoretical and computational models. Indeed, experimental facilities now provide ultrafast proton \cite{schaeffer2023proton}, neutron \cite{strobl2009advances}, and x-ray \cite{kozioziemski2023x} imaging capabilities, typically in the form of radiographic projections to characterize RMI behavior  \cite{endo1995dynamic,Serino24,zhai2018review,yager2022studying,prime2016using,do2022high,si2015experimental,zhai2018review}.
 While experimental RMI growth rates have been previously obtained in both planar and cylindrical geometries~\cite{rupert1992shock,brouillette2002richtmyer,zhou2021rayleigh,leinov2009experimental,holmes1999richtmyer,zhou2017rayleigh,zhang1998numerical}, validation of growth rates in spherical convergent geometries in which a reflected shock impacts a perturbed surface has not yet been achieved. This is in part due to the impact of noise/scatter in the radiographic images. The presence of scatter and noise observed in a typical radiograph, as demonstrated in Figure~\ref{fig:sampledensitysingle}, does not enable an accurate determination of the peak and troughs.

Consequently, characterization of the initial conditions responsible for the instability as well as material properties characterizing the growth rates demands new techniques to solve this inverse problem.  One such approach to resolve this difficulty has recently been proposed,\cite{hossain2022high} which utilizes the robust features of the outgoing shock to characterize the growth rates of the instability via a density reconstruction of the radiographic images. The success of this methodology is founded in the fact that the outgoing shock encodes sufficient information to enable machine learning techniques to learn a mapping between a sequence of outgoing shocks and the corresponding density fields. In this work we propose to utilize the outgoing shock as well as outer material edge as robust features to enable parameter estimation and estimation of the initial conditions. Using the recovered parameters in a hydrodynamic solver we then recover thermodynamically and hydrodynamically consistent density fields.


\subsection{A machine learning (ML)-based inverse mapping based on shock and outer edge features}

Here we develop a framework to construct machine-learning-based inverse mappings directly from experimental radiographic images to the underlying physical parameters and initial perturbations in ICF settings systems. 
Extraction of the parameters and initial perturbations 
enable subsequent use in a hydrodynamics solver to obtain
density fields, shocks, and material interfaces that satisfy \emph{hydrodynamic and thermodynamic consistency}
to numerical accuracy.
We assume a governing physical model given by the compressible Euler equations, dropping higher-order effects of radiation for ease of testing and development, analogous to recent work \cite{bello2020matrix}, and we assume unknown initial conditions and material properties for the governing Euler equations. 
As such, we build upon previous work that demonstrates the ability to learn parameters and density fields from a sequence of radiographic features for diagnosing asymmetries in the drive of ICF capsules by utilizing an inert gas as a surrogate for the D-T fuel to enable extraction of the outgoing shock~\cite{Serino23,Serino24,LA-UR-23-25917}. 
We  note that recent work at the National Ignition Facility has been performed using a silicon dopant ~\cite{le2014observation}. 
Furthermore, by subsequent design optimization to minimize asymmetries, improved neutron yields can be realized. 
Our specific problem description is one in which an ICF-like shell is imploded with an initially perturbed surface as depicted in \Cref{fig:initialcondition}. Upon collapse, the generation of a shock forms on axis and subsequently rebounds, interacting with the perturbed surface of the shell and initiates a RM instability. We posit that the outer material edge, outgoing shock, and their evolution \emph{encode} sufficient information from the instability to identify the underlying simplified Mie-Gr\"uneisenn EOS parameters and structure of the initial shell perturbation and velocity. Moreover, the shock and edge profiles are some of the few robust and identifiable features in dynamic radiographic images of HEDP experiments, which in general are subject to noise and scattering effects, and correspond to a projected areal mass rather than a primary hydrodynamic variable such as density. Thus broadly, we aim to calibrate material models and initial conditions to be consistent with the outgoing shock and edge profile, a problem of data-assimilation \cite{asch2016data}. There are many approaches to data assimilation, and here we review some of the prominent techniques, before concluding with our specific machine-learning-based approach. 

Variational data-assimilation minimizes a cost function comparing a forward model/simulation with experimental data \cite{asch2016data}. Although the field of data assimilation often uses emulators or surrogate models for forward evolution, when considering the full ``high-order'' forward model, variational data assimilation is a subclass of the broader mathematical fields of optimal control (unknown parameters/models) and inverse problems (unknown initial state) based on variational principles \cite{smith1998variational}. When the governing equations are dynamic partial differential equations (PDEs), each of these are PDE-constrained optimization problems. Indeed for problems related to ICF,
we believe it is critical to incorporate a high-order representation of the physics in a forward model, rather than working exclusively with some form of surrogate. As mentioned previously, shock interface and evolution provide some of the most robust and predictive information available, and surrogate or machine learning models that can accurately capture and track nonlinearly evolving and interacting shocks, let alone the formation of instabilities, remains a largely open question, e.g., see \cite{cai2022least}. 

PDE-constrained optimization is indeed a rigorous approach to inverse and optimal control problems, but is also very challenging computationally in terms of cost and memory \cite{biegler2003large}. Each gradient descent iteration requires a full forward and adjoint solution of the underlying PDE, in addition to storing a full physical solution to linearize about at \emph{every} time point simulated, in order to compute a gradient in the adjoint pass; additional difficulties arise in maintaining geometric structure \cite{tran2024properties}. This has led to significant recent interest in machine-learning based approaches to solving inverse and optimal control problems in computational physics. Perhaps the most popular are so-called physics informed neural networks (PINNs) \cite{raissi2019physics}, where the differentials of the underlying PDE are evaluated directly within a neural network via automatic differentiation. Then in training the PINN model, the forward and adjoint pass of classical gradient descent can be evaluated relatively cheaply purely based on automatic differentiation capabilities inherent to NNs. PINNs and many variations thereof have shown significant success in computational physics, particularly for optimal control and inverse problems, where the computational cost and required coding infrastructure are significantly less than a traditional adjoint-based optimization \cite{cuomo2022scientific}. 

{The emergence of machine learning has has brought about increased attention to data-driven approaches in ICF, high energy density, and plasma
physics research.  
Indeed, machine learning has been used to examine 
performance and sensitivity of ICF implosions~\cite{icfgaffney24}. Furthermore, Bayesian methods~\cite{icfgaffney13, icfhatfield20, icfwang24} have
been used with relatively simple physics models to infer parameters in
implosion experiments, such as laser-driven opacity and equation-of state from 
experiments. For an introduction and review of
current work, we recommend the review paper by Knapp and Lewis~\cite{knapplewis23}.
}

However, there is an additional major hurdle to overcome in calibrating models based on experimental radiographic data obtained from dynamic imaging experiments.  That is, these experiments  typically do not provide direct data on primary physical variables. Instead, the observations are images formed via both the primary signal as well as the scattered radiation signal along with the noise of the radiographic system and characteristics of the detection system. Indeed, extraction of the primary state variable, i.e. density, of the time-series of 2d noisy radiographic projections continues to be a challenge \cite{Serino24,hossain2022high,gautam2024learning}. Consequently, the computation of gradients based on radiographs, or extracted features and material interfaces as used here, necessary for PDE-constrained optimization or PINNs-like ML models immediately precludes the direct application of these methods.

Consequently, in this work we develop machine learning architectures that directly take in a low-dimensional time-series of radiographic images, extract the outgoing shock and outer material edge profiles, and estimate the corresponding EOS parameters as well as initial shell conditions. An additional benefit of this approach is that the estimated parameters can then be plugged back into a hydrodynamics code to yield physically admissible density field reconstructions of the radiographic images. Such physical consistency is lacking in virtually all other reconstruction techniques currently used. 
The generation of data is discussed in \Cref{sec:DataGeneration} and the machine learning (ML)-based parameter estimation pipeline is introduced in \Cref{sec:ArchitectureDescription}. Using numerical simulations, we then demonstrate in \Cref{sec:results} that our ML architecture is able to successfully recover EOS material parameters, initial shell profiles, and velocity to reproduce the observed shock and edge features, demonstrating that the time series of shock profiles does indeed encode sufficient information to infer these parameters with high accuracy. These parameters are then demonstrated to yield accurate and physically admissible feature and density reconstructions by evaluating the forward hydrodynamics code using predicted parameters. Moreover, in numerical simulation of HEDP, EOS is typically represented via underlying model assumptions or tabulated data models. 
We demonstrate that 
an expansive parameter model is sufficient to represent unknown models, 
in the sense that we can 
use an ML architecture trained on an
underlying EOS model, $M_1$, and output parameters of $M_1$ 
for a time-series of shock profiles generated with a \emph{different} underlying EOS model, $M_2$,
that lead to a consistent time-series of shock profiles as those originally generated based on $M_2$.
In this sense, our ML model is learning structure from the underlying physics, above a specific choice of EOS parameterization needed for numerical simulation. 




\section{Methods}

\label{sec:DataGeneration}

\subsection{Generation of density time series}

As a representative problem of a double shell ICF capsule implosion, we study shock propagation in a time-dependent pseudo-3D density profile, created by the implosion of a perturbed spherical metallic shell into a medium of gas. 
We utilize air in lieu of D-T in order to enable examination of the out-going shock which would otherwise be obscured by the burning plasma once the hot spot is formed.
We rely on the effects of a non-spherical perturbation and variation in initial velocity to provide distinct behavior in the late-time shock evolution. We assume
azimuthal symmetry, therefore the density at any time can be described in 2D cylindrical coordinates~$(r, z)$, but the solution remains 3-dimensional via the symmetry assumption.
Additionally, we focus on the Mie-Gr\"uneisen (MG) EOS model. We then generate a large set of data for variations in the inner shell perturbation, initial shell velocity, and EOS parameters, where each data sample contains a time series of the resulting hydrodynamic density field. 
Simulations are performed 
on a $440\times 440$
uniform Cartesian grid on a 
computational domain given by the quarter-plane $[0, L] \times [0, L]$, where 
$L=341$~$\mu$m.
The uniform grid cell size is $\Delta r = \Delta z = \frac{L}{440}$.
The metallic shell is made of Tantalum and its density is initially uniform at a value of 16.65~g/cc.

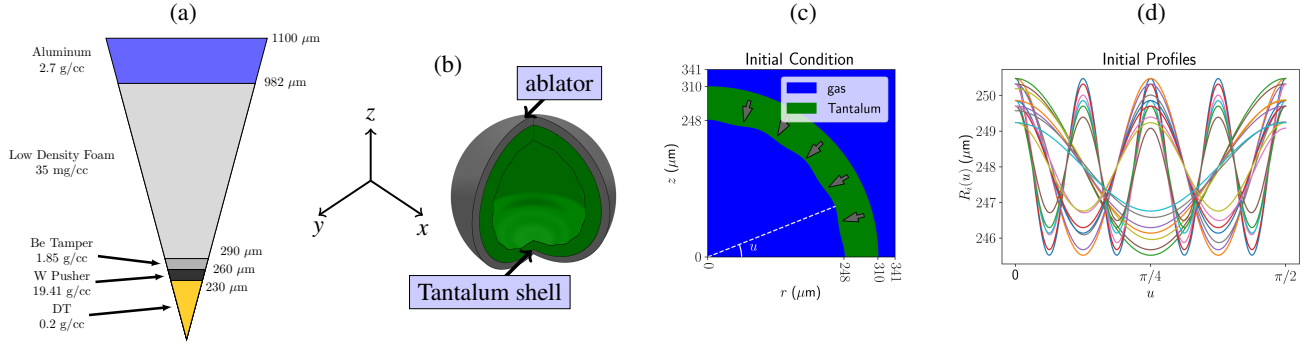
\begin{figure*}[htbp]
  \centering
  \resizebox{\textwidth}{!}{%
  \begin{tikzpicture}
  \useasboundingbox (-4,0.0) rectangle (15,4.75);  

  \draw(-1.8, -.5) node[anchor=south] {\trimh{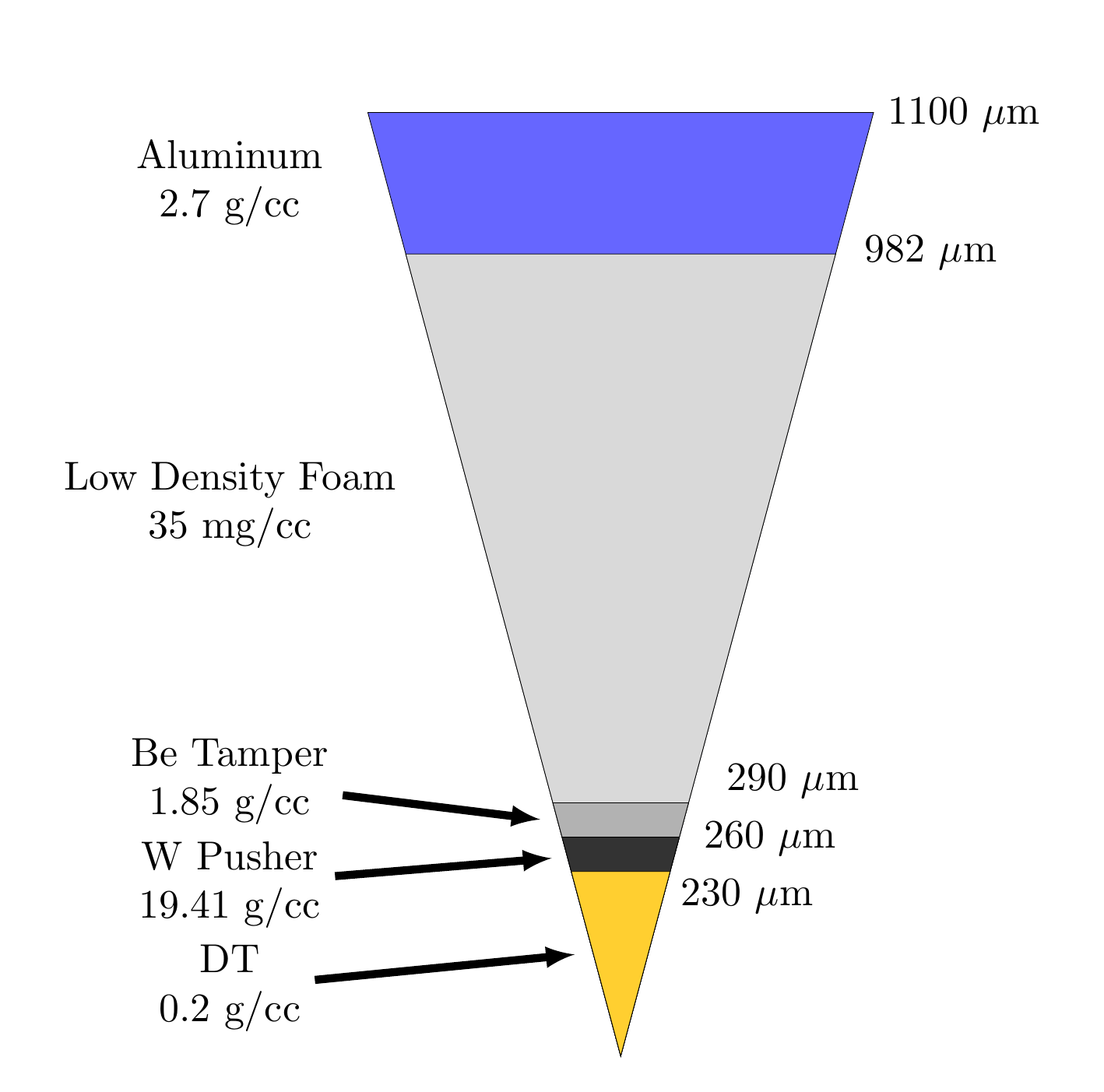}{5cm}{0}{0}{0}{0}};
  \draw(-1.4, 4.5) node[align=center] {(a)};

   \draw(3.5,0) node[anchor=south] {\trimh{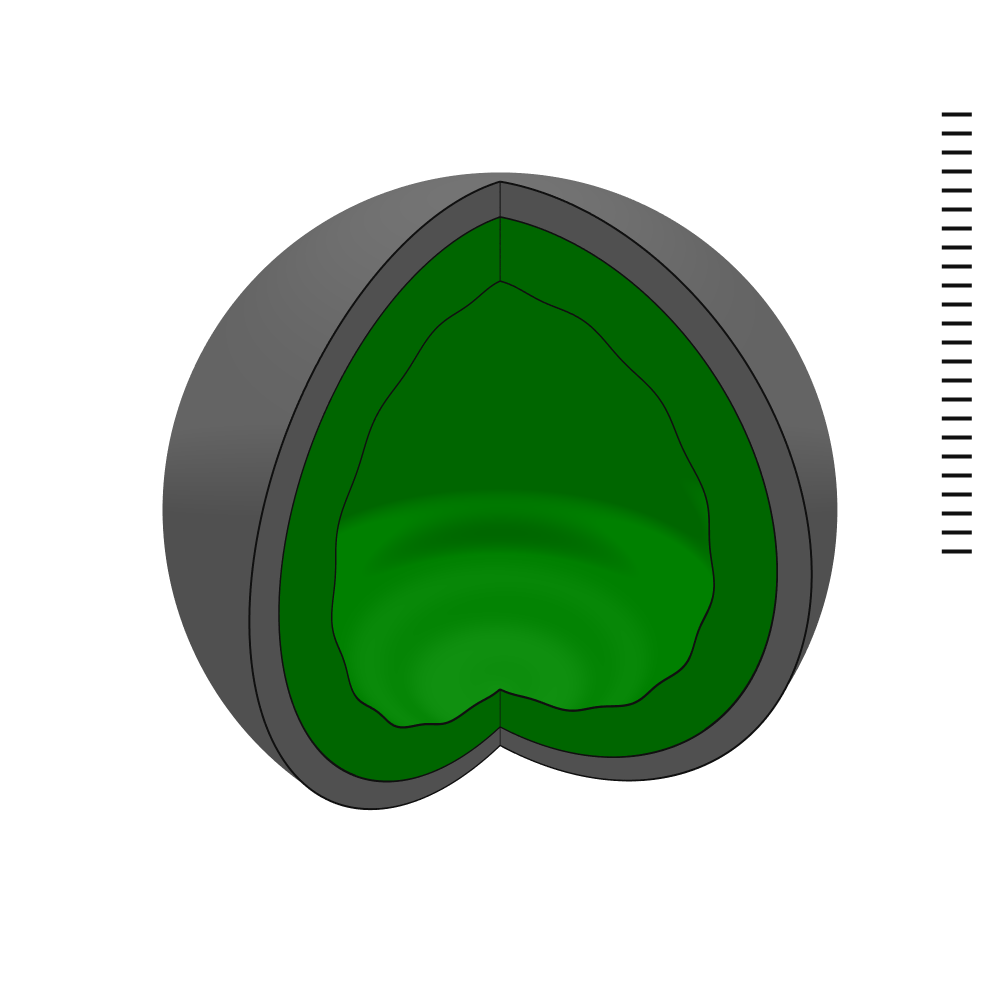}{3.5 cm}{.1}{.1}{0}{0}};
\def\xo{.54}
\def\yo{2.1};
  \draw[->, thick] (.75+\xo, 0+\yo) -- (1.5+\xo, -.5+\yo) node[yshift=-.2cm] {$x$};
  \draw[->, thick] (.75+\xo, 0+\yo) -- (.75+\xo, .75+\yo) node[yshift=.2cm] {$z$};
  \draw[->, thick] (.75+\xo, 0+\yo) -- (0.0+\xo, -.5+\yo) node[yshift=-.2cm] {$y$};
  
  \draw(2.4, 3.75) node[align=center] {(b)};

   \draw(10.25,0) node[anchor=south east] {\trimh{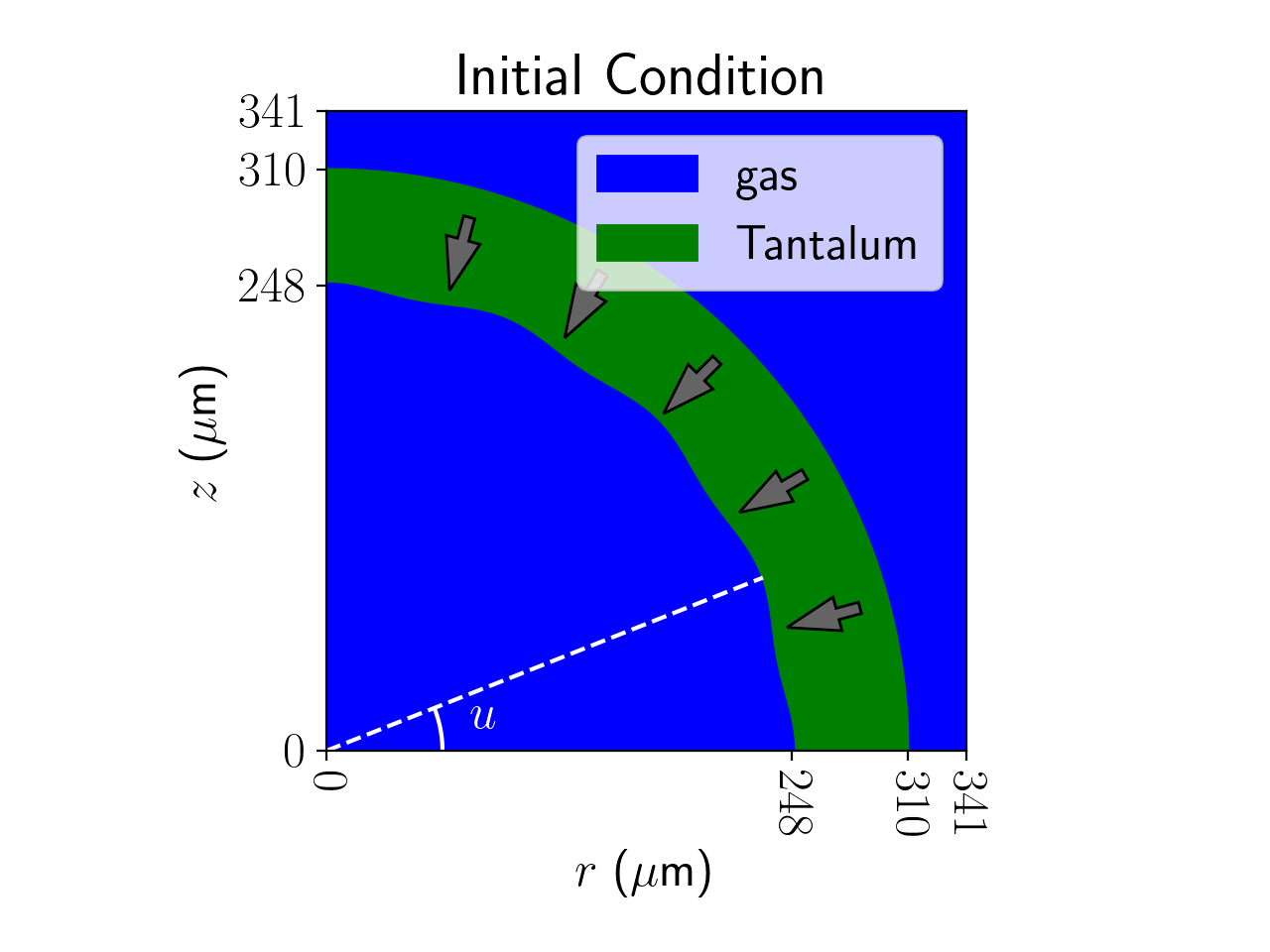}{4 cm}{.05}{.0}{0}{0}};

     \draw(7.5, 4.5) node[align=center] {(c)};

   \draw(15,0) node[anchor=south east] {\trimh{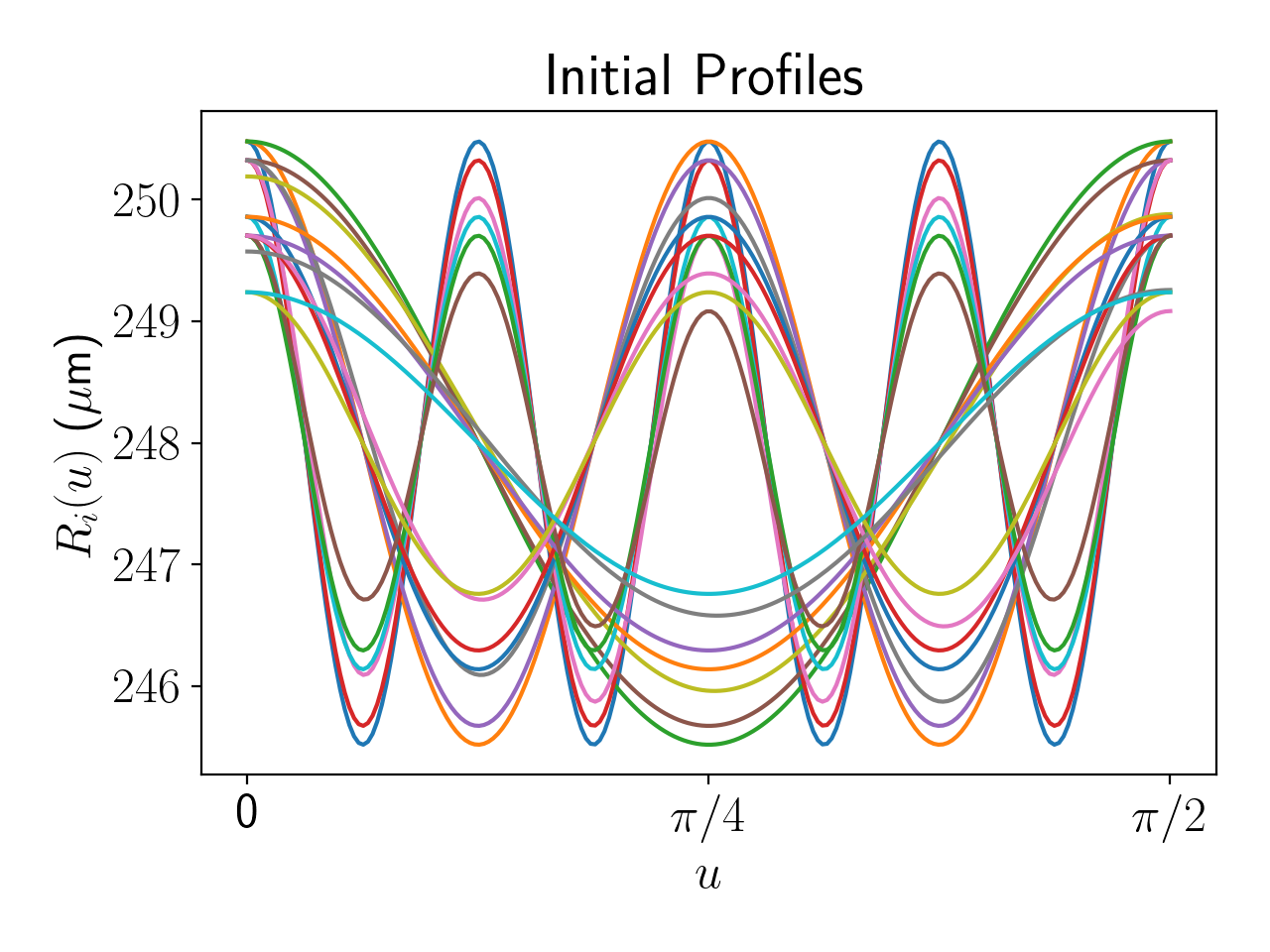}{4 cm}{.0}{.0}{0}{0}};
        \draw(12.5, 4.5) node[align=center] {(d)};

   \coordinate (A) at (4, 3.5);
   \coordinate (B) at (3, .5);
   \draw[->, line width=1.5 pt] (A) -- (3.5, 3);
   \draw[->, line width=1.5 pt] (B) -- (3.6, 1.1);        
   \draw (A) node[draw, fill=blue!20] {ablator};
   \draw (B) node[draw, fill=blue!20] {Tantalum shell};
  \end{tikzpicture}  
  }
  \caption{
  (a): Example double shell capsule specification based on the 1.06 MJ yield design from Ref.~\cite{merritt19}.
  (b): 3D mock-up of a shell with a
  perturbation on the interior surface. 
  (c): projection of the shell onto $(r, z)$ 
  coordinates. The inner radius is parameterized by
  the angle, $u$ between the white dotted line and
  the $r$ axis.
  (d): Plot of the 20 separate profiles for 
  radius of the perturbed inner surface verses angle $u$. 
  }
  \label{fig:initialcondition}
\end{figure*}

The shell perturbations are specified by adding harmonic perturbations to the inner surface of a spherical Tantalum shell, which can be described as the set of 
coordinates $(r_{\rm in}(u), z_{\rm in}(u))$ satisfying 
$r_{\rm in}(u)^2 + z_{\rm in}(u)^2 = \left(R_{\rm in} 
+ \sum_{k=1}^{8} F_k \cos(2k u)\right)^2$,
%
where $u \in [0, \pi/2]$, ${R_{\rm in} = 248}$ $\mu$m, $F_k$, $k=1,\dots,8$, are coefficients of the perturbation corresponding to the $k^{\rm th}$ cosine harmonic.
The outer surface of the shell is a sphere with radius ${R_{\rm out} = 310}$ $\mu$m.
There are 20 different inner surface perturbation profiles considered in our dataset. The
corresponding coefficients are recorded in Table~\ref{tab:initialcoeffs}.
Figure~\ref{fig:initialcondition} presents an initial perturbation given to the interior shell.
As an initial condition, the shell is given a uniform implosion velocity, $v_{\rm impl}$,
in the direction of the origin to initiate an implosion.
We include 4 choices of implosion velocity in our dataset. 

The MG EOS~\cite{hertel98b} can be parameterized in analytical form as
\begin{align}
    p\left(\chi, T\right) 
    = \frac{\rho_0 c_s^2\chi\left(1 - \frac12\Gamma_0\chi\right)}
          {(1 - s_1\chi)^2}
    + \Gamma_0\rho_0 c_V (T - T_0),
\end{align}
where $\chi=1-\frac{\rho_0}{\rho}$,
$\rho_0$ and $T_0$ are the reference density and temperature, respectively, $c_s$ is the speed of sound, $\Gamma_0$ is the Gr\"uneisen parameter at the reference state, $s_1$ is the slope of the linear shock Hugoniot curve, and $c_V$ is the specific heat capacity at constant volume.
Out of these parameters, we keep the reference density $\rho_0$ fixed at 16.65~g/cc and the reference temperature $T_0$ fixed at 0.0253~eV, leaving EOS parameters $\{c_s,\, s_1,\, \Gamma_0,\, c_V\}$ as unknown. 
Table~\ref{tab:vary} presents the EOS parameter values we sample in generating our training data set. 

\begin{table}[htbp]
  \centering
  \resizebox{.6\columnwidth}{!}{%
  \begin{tabular}{|c|ccccc|} 
  \hline 
    Options  & 1   & 2  & 3 & 4 & 5 \\
  \hline
    $\Gamma_0$  & 1.6   & 1.7  & 1.76 & 1.568 & 1.472 \\
  \hline
    $s_1$  & 
    1.22 & 1.464  & 1.342 &  & \\
  \hline
    $c_s$ [m/s]  & 339000   & 372900  & 305100 & 355000 &  \\
  \hline
    $c_V$ $[{\rm erg}\;{\rm g}^{-1}\;{\rm eV}^{-1}]$  & $1.6\times 10^{10}$   & $1.76 \times 10^{10}$  & $1.44 \times 10^{10}$ &    &  \\
  \hline
  \end{tabular}
  }
  \caption{Matrix of parameter values used to develop the simulated dataset. All combinations of above parameters are used to simulate our data.
  }
  \label{tab:vary}
\end{table}

Altogether, the dataset realizes every unique parameter combination in a 6-dimensional parameter cube with $14,400$ total simulations. 
Each hydrodynamic simulation is comprised of 
density field snapshots 
at later times when the instability is present. 
We label these times as $n=0, 1, \dots, 40$.
An example of a density time series is shown in
Figure~\ref{fig:dyn_egs}a.
Once the shock propagating through the gas converges to the axis, a reflected shock from the axis then propagates outward and interacts with the perturbed inner Tantalum edge, creating a RMI. The topology of this interior evolves as depicted in Figure~\ref{fig:dyn_egs}b. The expanding shock proceeds to propagate into the
non-constant dynamic density background. 
We chose frames corresponding to the time instants at $n=25, 30, 35, 40$
to train the network in our studies. 

\subsection{Generation of Synthetic Radiographs}

Synthetic radiographs are produced at each time step
using the imaging model described in~\cite{Serino24}.
This model involves first obtaining the areal mass
for the gas and metal using the cone beam projection
provided by the ASTRA Toolbox~\cite{astra}.
The final transmission includes contamination from several noise terms,
which contains correlated blur, scatter, and a Poisson noise field.
Figure~\ref{fig:sampledensitysingle} shows an example of a density field at
time index 40 and a synthetic radiograph.

\subsection{Identification of shock and edge features}

One of the primary aspects of the dynamics is the 
evolution of the inner air-metal interface,
i.e., the growth of the instability. This is because the
passage of the incoming and outgoing shocks through this interface
renders it unstable to the RMI. Considering temporally evolving simulations, we
are interested in times when the instability on this interface has
permitted the growth of perturbations to the extent that the inner
air-metal interface displays significant asymmetry. 
As such, we assume
that the interface as identified by a feature extraction procedure
is not robust due to sensitivities with respect to the chosen imaging plane. 
This is in contrast to the shock and
outer edge features that we assume are robust. Nevertheless, because of the shock's
passage across the unstable inner air-metal interface, we expect the
stably evolving shock to be imprinted with a set of
perturbations that can be reliably identified in a noisy radiograph~\cite{Serino24}. 

Shock and edge features are extracted at each time for each sequence of density fields.
Our feature extraction algorithm consists of two stages: (i) using the maximal gradient to detect computational cells where the features are present; and (ii) subpixel feature extraction.
For subpixel feature detection, we use the partial-area algorithm~\cite{trujillo-pino2013}, generalized to quadratic density functions assumed on both sides of the feature.
Specifically, we examine the neighborhood of every pixel where the feature is present, assuming that the density on both sides of the discontinuity may be described by a function which is quadratic in both variables.
Following the original algorithm~\cite{trujillo-pino2013}, we then fit for the shape of the feature with the constraint that the mass along the stripes in the neighborhood of the feature matches the integrated density.
The result is a parametric representation of the shock and edge as a function of polar angle~\cite{hossain2022high}.
We compressed these shock and outer edge features into a low-dimensional 
representation in terms of cosine harmonic coefficients,
$r^{(i)}(\theta) = \sum_{j=0}^{N^{(i)}} F^{(i)}_j \cos(2 j \theta),$
%
for $i={\rm shock},{\rm edge}$. We found
that $N^{(\rm shock)}=8$ and $N^{(\rm edge)}=5$
can represent the shock and edge features with sufficient 
accuracy across the dataset.

\subsection{Parameter estimation machine learning pipeline}
\label{sec:ArchitectureDescription}

Our ML-based 
parameter estimation pipeline is composed of two architectures, a 
radiograph-to-features network (R2FNet) 
and a features-to-parameters (F2PNet) network, 
described respectively in Sections~\ref{sec:r2f} and~\ref{sec:f2p}.
R2FNet and F2PNet are trained separately and later combined for testing.
F2PNet also consists of a forward model,
which is a surrogate for the parameters-to-features
mapping.
In Section~\ref{sec:results}, we evaluate the model's ability to recover parameters from radiographs 
through examining self consistency with respect to features
produced by both the surrogate and true forward model.
We consider a time sequence of $n\in\{25, 30, 35, 40\}$,
which corresponds to times where the outgoing 
shock is fully formed in the metal and the 
RMI has entered a linear growth phase.
The dataset consisting of 
triplets of radiographs, features, and parameters
is randomly partitioned into training, validation,
and testing sets corresponding to 
80\%, 10\%,
and 10\% 
of the data, respectively.

\section{Results}\label{sec:results}


\subsection{Model Performance on Testing Set}

\begin{figure*}[!htb]
    \centering
    \resizebox{\textwidth}{!}{%
  \begin{tikzpicture}
  \useasboundingbox (0,-0.5) rectangle (15,12);  

  \draw(7.5, 4.5) node[anchor=south] {\trimwb{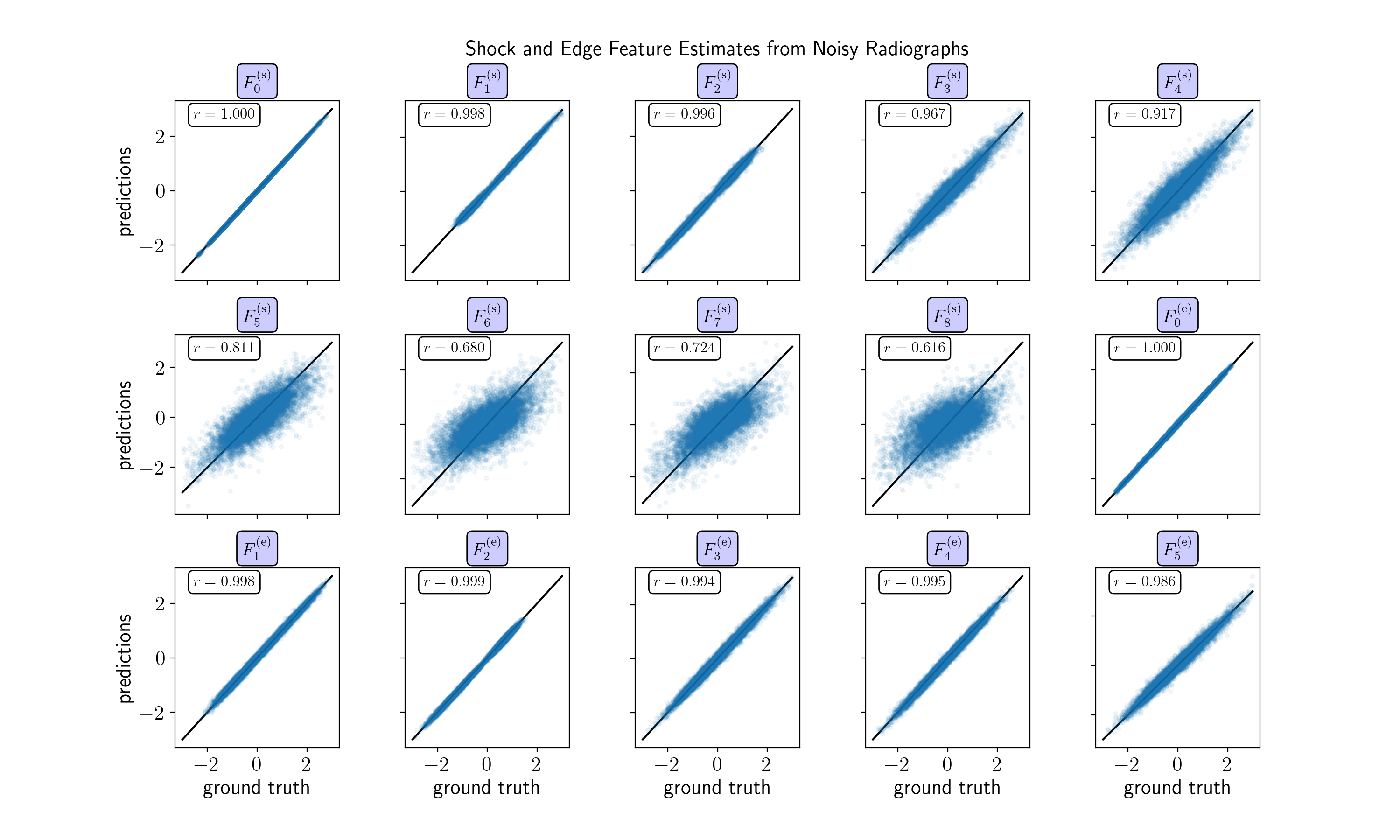}{11cm}{.07}{.09}{.02}{.03}};

  \draw(7.5, -1) node[anchor=south] {\trimwb{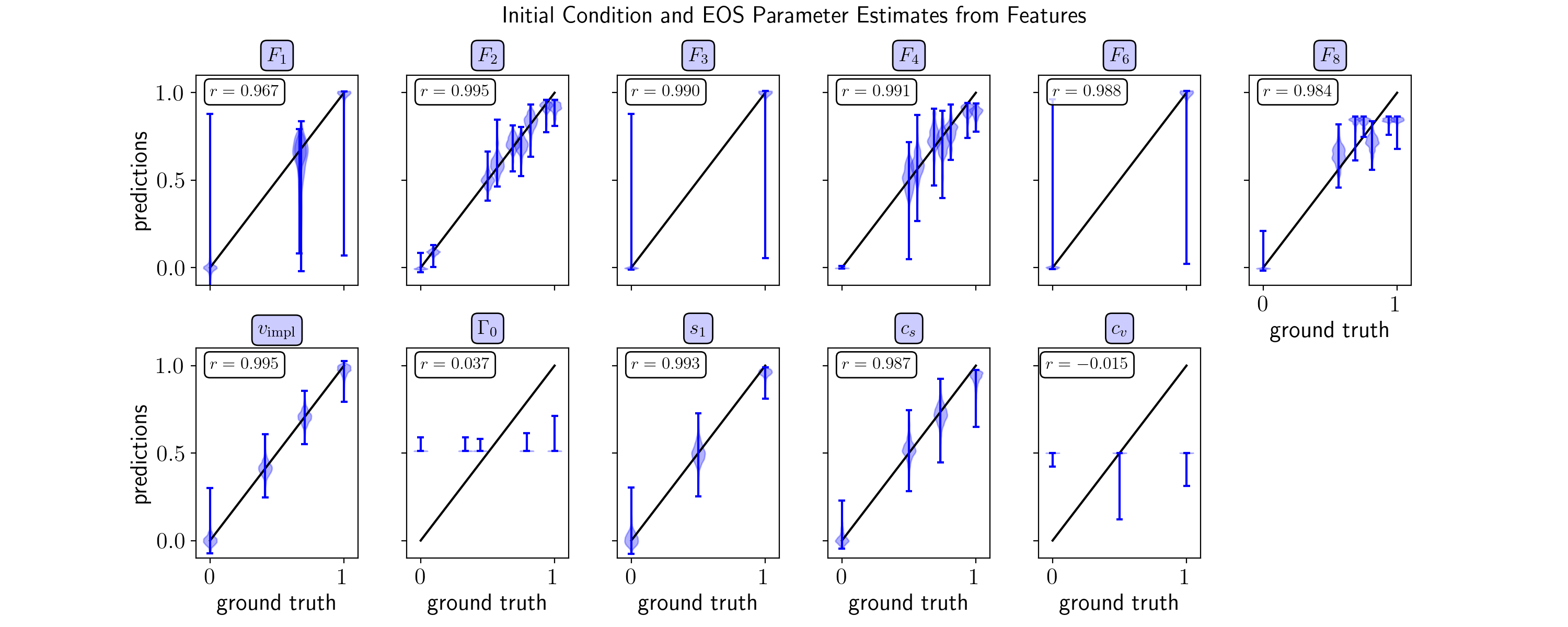}{11cm}{.07}{.09}{.0}{0}};
    \draw(1.7, 8.5) node[] {(a)};
    \draw(1.7, 1.8) node[] {(b)};

\bogus{
\draw(15.2, 5.75) node[anchor=south east, align=center, text width=4.25cm, scale=.8] {
\resizebox{1\columnwidth}{!}{%
    \begin{tabular}{|c|cc|}
    \hline
feature & mean & std. \\
\hline
$F_0^{\rm (s)}$ &   1.04e+02 &   1.86e+01 \\
$F_1^{\rm (s)}$ &   7.12e-01 &   7.93e-01 \\
$F_2^{\rm (s)}$ &  -1.73e-01 &   4.26e-01 \\
$F_3^{\rm (s)}$ &   7.39e-03 &   1.19e-01 \\
$F_4^{\rm (s)}$ &  -2.89e-02 &   7.27e-02 \\
$F_5^{\rm (s)}$ &  -6.22e-03 &   4.13e-02 \\
$F_6^{\rm (s)}$ &   2.23e-02 &   3.63e-02 \\
$F_7^{\rm (s)}$ &   3.30e-03 &   3.37e-02 \\
$F_8^{\rm (s)}$ &  -2.06e-02 &   3.30e-02 \\
$F_0^{\rm (e)}$ &   3.07e+02 &   9.14e-01 \\
$F_1^{\rm (e)}$ &   1.28e-01 &   1.15e-01 \\
$F_2^{\rm (e)}$ &  -2.44e-01 &   2.00e-01 \\
$F_3^{\rm (e)}$ &   2.77e-02 &   5.35e-02 \\
$F_4^{\rm (e)}$ &  -4.42e-02 &   6.90e-02 \\
$F_5^{\rm (e)}$ &   3.60e-03 &   3.67e-02 \\
\hline
    \end{tabular}}

};

\draw(15.2, 0) node[anchor=south east, align=center, text width=4.25cm, scale=.8] {
\resizebox{1\columnwidth}{!}{%
    \begin{tabular}{|c|cc|}
    \hline
parameter & min & max \\
\hline
$F_1$ & 0.0 &   3.00e-01 \\
$F_2$ & 0.0 &   3.20e+00 \\
$F_3$ & 0.0 &   1.00e-01 \\
$F_4$ & 0.0 &   3.20e+00 \\
$F_6$ & 0.0 &   1.00e-01 \\
$F_8$ & 0.0 &   3.20e+00 \\
$\Gamma_0$ & 1.47 & 1.76 \\
$s_1$ & 1.22 & 1.46 \\
$c_s$ & 3.05e+05 & 3.73e+05 \\
$c_v$ & 1.44e+10 & 1.76e+10 \\
PTW & 6.50e+10 & 6.50e+11 \\
\hline
    \end{tabular}}

};
}
    
  \end{tikzpicture}
}
    \bogus{
    \includegraphics[trim=0cm 0cm 0cm 0.0cm, clip, width=.8\linewidth]{paper_figs/r2f_results.png}    
    \includegraphics[trim=2cm .25cm 3cm 0.0cm, clip, width=.85\linewidth]{figs/line_params.png}
    }
    \caption{
    Prediction performance of the parameter estimation pipeline
    on the testing set.
    (a):
    Scatter plots depicting the agreement between the scaled 
    features predicted using R2FNet on the test set
    of noisy radiographs (vertical axis) and the
    corresponding scaled ground
    truth features (horizontal) axis in the testing set.
    (b):
    Line plots depicting the agreement between the 
    scaled parameters predicted using F2PNet on the features
    predicted from R2FNet
    (violin plots representing the data 
    range and distribution) and scaled ground 
    truth parameters (diagonal lines) in the testing set. The correlation coefficients are 
    shown in their respective plots. }
    \label{fig:lines}
\end{figure*}

After training R2PNet and F2PNet, parameter estimations
were performed on each data point in the testing set.
R2PNet is used to predict shock and edge features for each 
radiograph.
These predicted features are then inputted into the decoder of 
F2PNet for 25 different realizations of the generator
to produce a distribution of parameter estimates.
The feature and parameter predictions are illustrated in
Figure~\ref{fig:lines}.
The horizontal axes correspond to ground truth values and 
the vertical axes correspond to the predictions.
In the case of the parameter predictions, 
violin plots are shown on the vertical axis corresponding to the 
distribution of network predictions.
Correlation coefficients are shown in 
each
of the plots.

As seen by the plots, R2FNet is able to predict the 
edge and the lower order harmonics of the 
shock (e.g. $F_0$ through $F_4$) 
with high correlation, while correlation
is degraded for higher order harmonics of the shock (e.g. $F_1$ through $F_8$).
Despite this, F2PNet is able to predict all parameters
except for $\Gamma_0$ and $c_v$
with high correlation.
This demonstrates that $\Gamma_0$ and $c_v$ are likely insensitive 
parameters for late-time shock and edge features and the rest of the 
parameters can be predicted with reasonable accuracy.
The violin plots show that the range of estimates can be highly variable,
however the distributions seem to be closely centered around the 
diagonal line denoting ground truth.
Note that $F_0, F_5,$ and $F_7$ are omitted since there is no variance in these parameters in the dataset.

\begin{figure*}[!htb]
    \centering   
    \includegraphics[trim=2cm 1cm 2.5cm 3cm, clip,width=.8\linewidth]{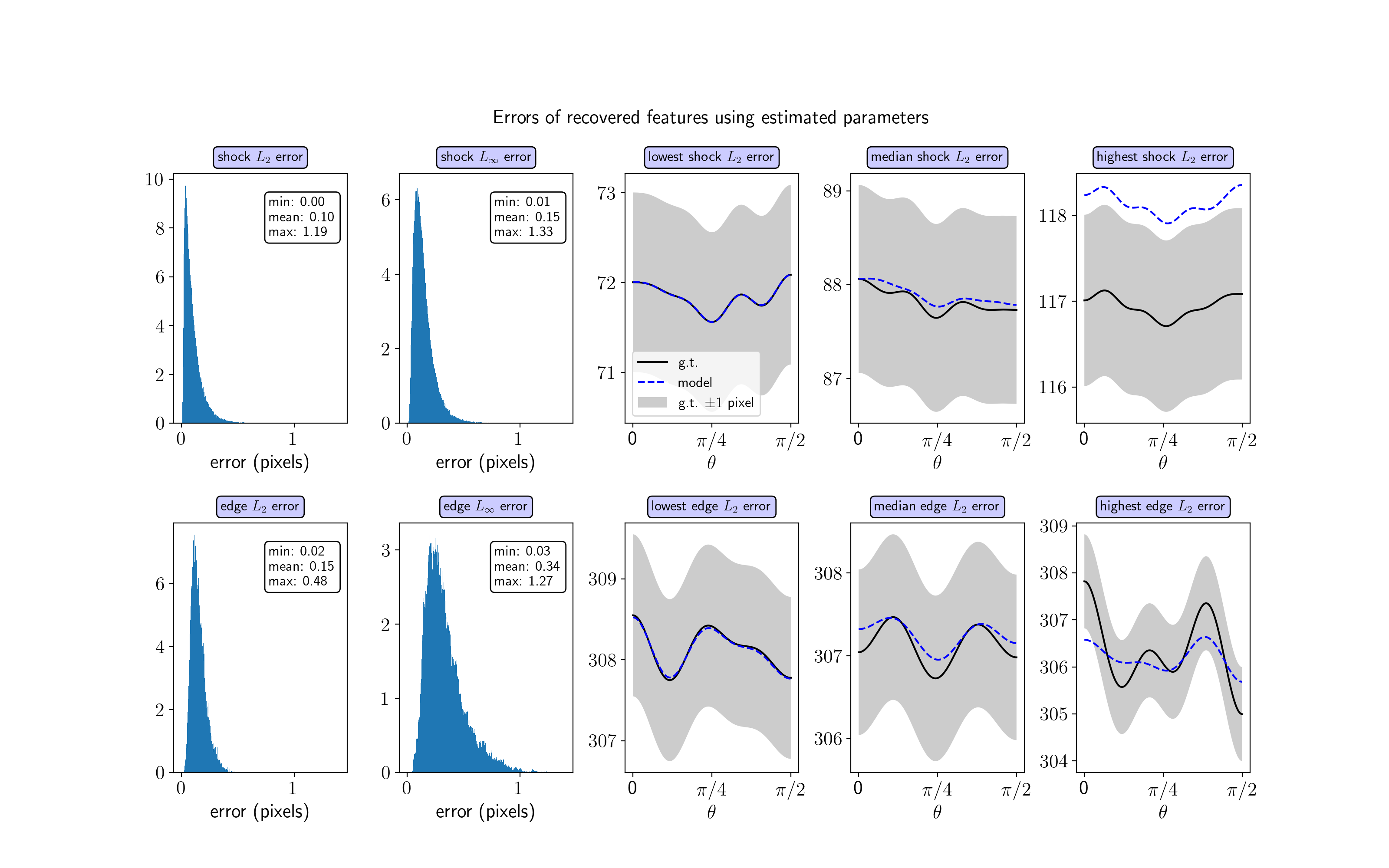}
    \caption{
    Errors of reconstructed shock and edge features reconstructed using the forward model on parameters estimates from the decoder on the testing set. The top row corresponds to shock features and the bottom row
    corresponds to edge features.
    The first two columns show histograms of the $L_2$ and $L_\infty$ errors,
    respectively, and the last three columns show examples of reconstructed
    shock and edge features (dotted blue line) representing the extreme cases of the
    distributions superimposed over the ground truth
    feature (black line) and an error bar of $\pm$1 pixels 
    (gray region).
    }
    \label{fig:ae_features}
\end{figure*}

We now evaluate the capability of the trained forward model
along with the self-consistency of the network.
Feature predictions are obtained by inputting the 
parameter estimates from the decoder on the testing set
into the forward model.
Figure~\ref{fig:ae_features}
shows histograms of the $L_2$ and $L_\infty$ errors for the shock and edge curves resulting from the feature predictions.
The majority of the reconstructions are accurate to within one pixel of error. 
Figure~\ref{fig:ae_features} shows example 
feature reconstructions corresponding to the
best, median, and worst $L_2$ errors for both the
shock and edge.

\subsection{Model Sensitivity Studies}

We now consider the effect of the training set size on predictive capability.
Models are trained with various training set sizes
ranging from 10\% to 70\%. 
After training, correlation coefficients are computed 
for each parameter and shown in Table~\ref{tab:data_size}.
Monotonic improvement is generally observed across all parameters
with $c_v$ as an exception. 
It is clear that there is significant degradation in performance when the sample size is too small.
Additionally, we observe that 
the degradation is dependent on the parameter.
We remark that, relative to other parameters, $F_2, F_4$, and $F_8,$ are learned with reasonable skill with the smallest training set; for harmonics, this makes sense given that $F_2, F_4$, and $F_8$ are more densely sampled compared to other parameters (see Figure~\ref{fig:lines}). 
Note that even at 30\% of the available training data, the correlation coefficients for many of the parameters do not drop significantly.
These insights are important for our future work on 
3D problems that are not assumed to be symmetric, 
since the data generation is significantly more computationally expensive.

\begin{table}[htb]
    \centering
    \large{Training Set Size Performance Study}

    \smallskip
    
    \resizebox{.8\columnwidth}{!}{%
    \begin{tabular}{|c|ccccccccccc|}
    \hline
$f_{N_d}$ & $F_1$ & $F_2$ & $F_3$ & $F_4$ & $F_6$ & $F_8$ & $v_{\rm impl}$ & $\Gamma_0$ & $s_1$ & $c_s$ & $c_v$ \\
\hline
10\% & 0.575 & 0.954 & 0.607 & 0.900 & 0.595 & 0.880 & 0.536 & 0.007 & 0.604 & 0.631 & -0.009 \\
30\% & 0.866 & 0.976 & 0.893 & 0.956 & 0.903 & 0.972 & 0.914 & 0.011 & 0.942 & 0.899 & -0.016 \\
50\% & 0.924 & 0.996 & 0.938 & 0.966 & 0.927 & 0.987 & 0.940 & 0.076 & 0.974 & 0.941 & 0.000 \\
70\% & 0.950 & 0.978 & 0.961 & 0.972 & 0.953 & 0.995 & 0.973 & 0.252 & 0.988 & 0.976 & -0.023 \\
\bogus{
$f_{N_d}$ & $F_1$ & $F_2$ & $F_3$ & $F_4$ & $F_6$ & $F_8$ \\
\hline
10\% & 0.575 & 0.954 & 0.607 & 0.900 & 0.595 & 0.880  \\
30\% & 0.866 & 0.976 & 0.893 & 0.956 & 0.903 & 0.972  \\
50\% & 0.924 & 0.996 & 0.938 & 0.966 & 0.927 & 0.987  \\
70\% & 0.944 & 0.977 & 0.955 & 0.969 & 0.951 & 0.994  \\
\hline \hline
$f_{N_d}$ & $v_{\rm impl}$ & $\Gamma_0$ & $s_1$ & $c_s$ & $c_v$ & \\
\hline
10\% & 0.536 & 0.007 & 0.604 & 0.631 & -0.009 & \\
30\% & 0.914 & 0.011 & 0.942 & 0.899 & -0.016 &\\
50\% & 0.940 & 0.076 & 0.974 & 0.941 & 0.000 &\\
70\% & 0.971 & 0.252 & 0.987 & 0.975 & -0.024 &\\
}
\hline
    \end{tabular}
    }
    \caption{Correlation coefficients of parameter predictions
    evaluated on the testing set for a model trained 
    using different data-set fractions ($f_{N_d}$). 
    }
    \label{tab:data_size}
\end{table}

\begin{figure}[htb]
    \centering
\resizebox{\textwidth}{!}{%
  \begin{tikzpicture}
  \useasboundingbox (0,0) rectangle (15,4.25);  

  \draw(-1, .5) node[anchor=south west] {\trimw{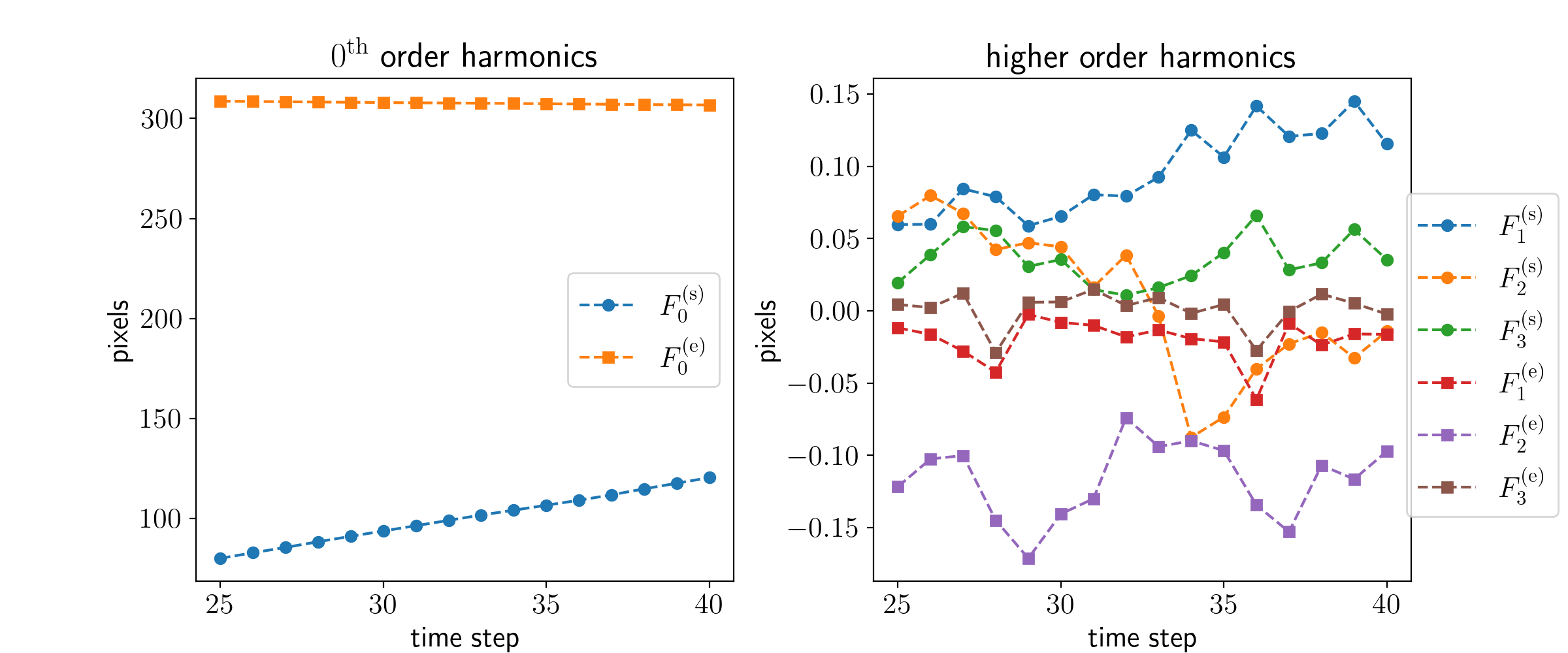}{9.4cm}{.0}{.0}{.0}{0}};

\draw(15, 2.5) node[anchor=south east, align=center, text width=6cm] {
\resizebox{1\columnwidth}{!}{%
    \begin{tabular}{|c|cccccc|}
    \hline
case & $F_1$ & $F_2$ & $F_3$ & $F_4$ & $F_6$ & $F_8$ \\
\hline
{\small all features} & 0.954 & 0.998 & 0.962 & 0.991 & 0.949 & 0.992 \\
{\small $F_0^{\rm (s)}$ only} & 0.344 & 0.842 & 0.402 & 0.663 & 0.460 & 0.747 \\
\hline    
    \end{tabular}
    }

};

\draw(15, 1) node[anchor=south east, align=center, text width=6cm] {
\resizebox{1\columnwidth}{!}{%
    \begin{tabular}{|c|ccccc|}
    \hline
case & $v_{\rm impl}$ & $\Gamma_0$ & $s_1$ & $c_s$ & $c_v$ \\
\hline
{\small all features} & 0.974 & 0.011 & 0.985 & 0.970 & -0.005 \\
{\small $F_0^{\rm (s)}$ only} & 0.929 & 0.146 & 0.980 & 0.916 & -0.010 \\
\hline    
    \end{tabular}
    }
};
    \draw(2.3, .25+0*4.75) node[] {(a)};
    \draw(6.25, .25+0*4.75) node[] {(b)};
    \draw(12, .25+0*4.75) node[] {(c)};
    \draw(12, 3.8) node[] {Harmonics Performance Study};

  \end{tikzpicture}
}

\bogus{
\includegraphics[width=.75\linewidth]{}
    
    \large{Harmonics Performance Study}

    \smallskip
        
    \resizebox{.9\columnwidth}{!}{%
    \begin{tabular}{|c|ccccccccccc|}
    \hline
case & $F_1$ & $F_2$ & $F_3$ & $F_4$ & $F_6$ & $F_8$ & $v_{\rm impl}$ & $\Gamma_0$ & $s_1$ & $c_s$ & $c_v$ \\
\hline
{\small all features} & 0.954 & 0.998 & 0.962 & 0.991 & 0.949 & 0.992 & 0.974 & 0.011 & 0.985 & 0.970 & -0.005 \\
{\small 0$^{\rm th}$ shock harmonic only} & 0.344 & 0.842 & 0.402 & 0.663 & 0.460 & 0.747 & 0.929 & 0.146 & 0.980 & 0.916 & -0.010 \\
\hline    
    \end{tabular}
    }
    }
    \caption{
    (a): Line plot illustrating the dynamics of the $0^{\rm th}$ order harmonic
    coefficients
    of the shock, $F_0^{\rm (s)}$, and edge, $F_1^{\rm (e)}$ for 
    a selected example in the dataset.
    (b): Corresponding line plot illustrating the dynamics of the 
    next three higher
    order harmonic coefficients of the shock and edge.
    (c): Correlation coefficients of parameter predictions
    evaluated on the testing set for a model trained 
    using the entire set of shock and edge features 
    compared to a model trained only using the 
    0$^{\rm th}$ harmonic of the shock.}
    \label{tab:0th_harm}
\end{figure}

In Figure~\ref{tab:0th_harm}c, we consider a study of comparing a baseline trained network
to a network trained only using the $0^{\rm th}$ harmonic of the shock feature ($F_0^{(s)}$).
Despite the significant decrease in the feature dimension, $v_{\rm impl}$ and many of the 
EOS parameters, including $s_1$, and $c_s$, 
can be recovered 
accurately. In contrast, poor accuracy is achieved in prediction of initial perturbation harmonic coefficients, which is intuitive given the geometric relation of initial perturbation and perturbations to the outgoing shock profile.  
It is also instructive to examine the temporal evolution of the magnitude of the harmonics, as seen for an example trajectory
in Figure~\ref{tab:0th_harm}a-\ref{tab:0th_harm}b.  
Although the $0^{\rm th}$ harmonic of the shock is growing, there is no apparent growth trend in the higher harmonics of the shock.
Therefore, the relative information content 
provided by the higher order harmonics 
diminishes in time.

\bogus{
\begin{figure}[ht]
    \centering
    \begin{subfigure}{.49\linewidth}
    \centering
        \includegraphics[width=\linewidth]{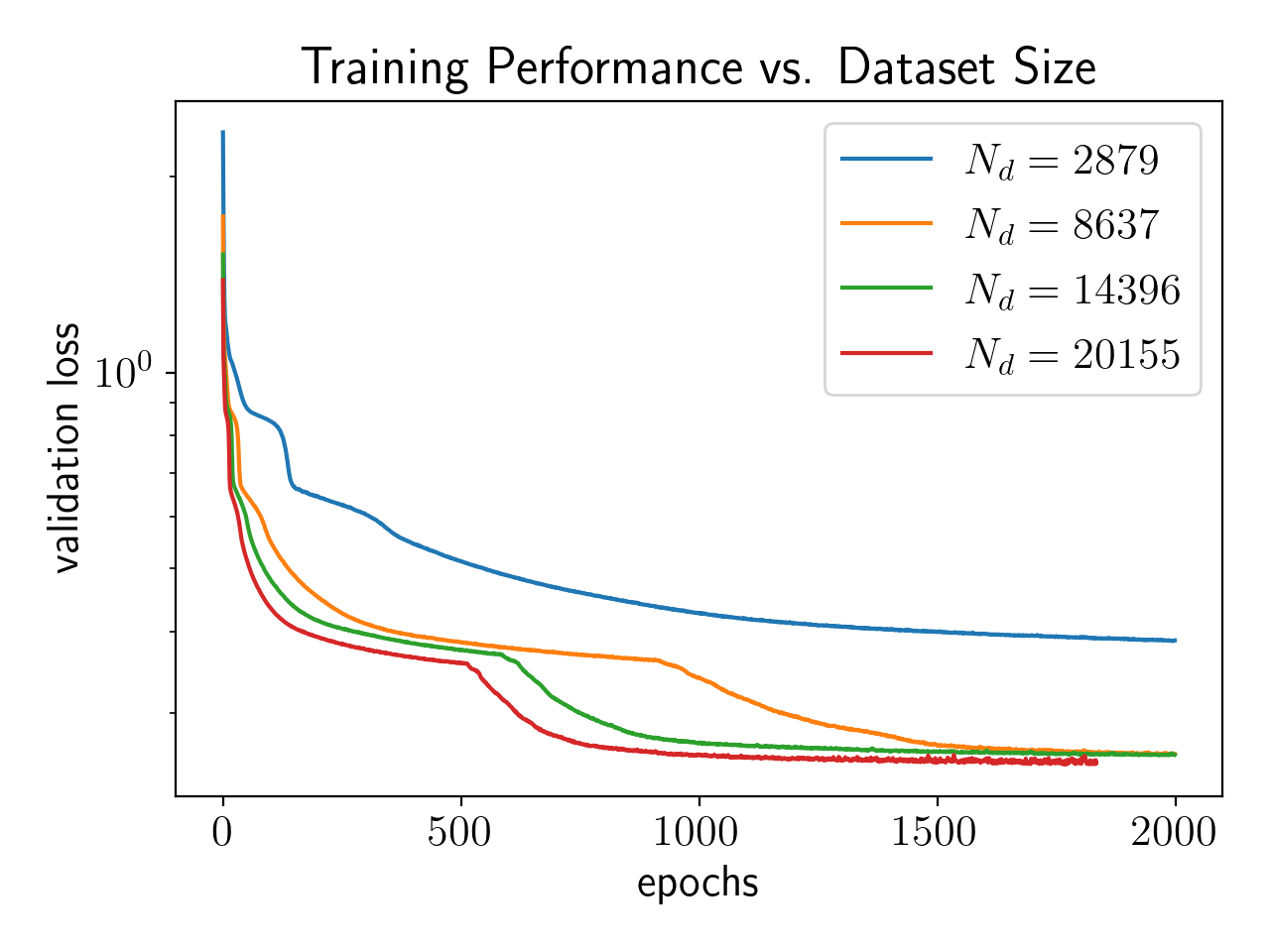}
        \caption{\label{fig:comp_a}}
    \end{subfigure}
    \begin{subfigure}{.49\linewidth}
    \centering
        \includegraphics[width=\linewidth]{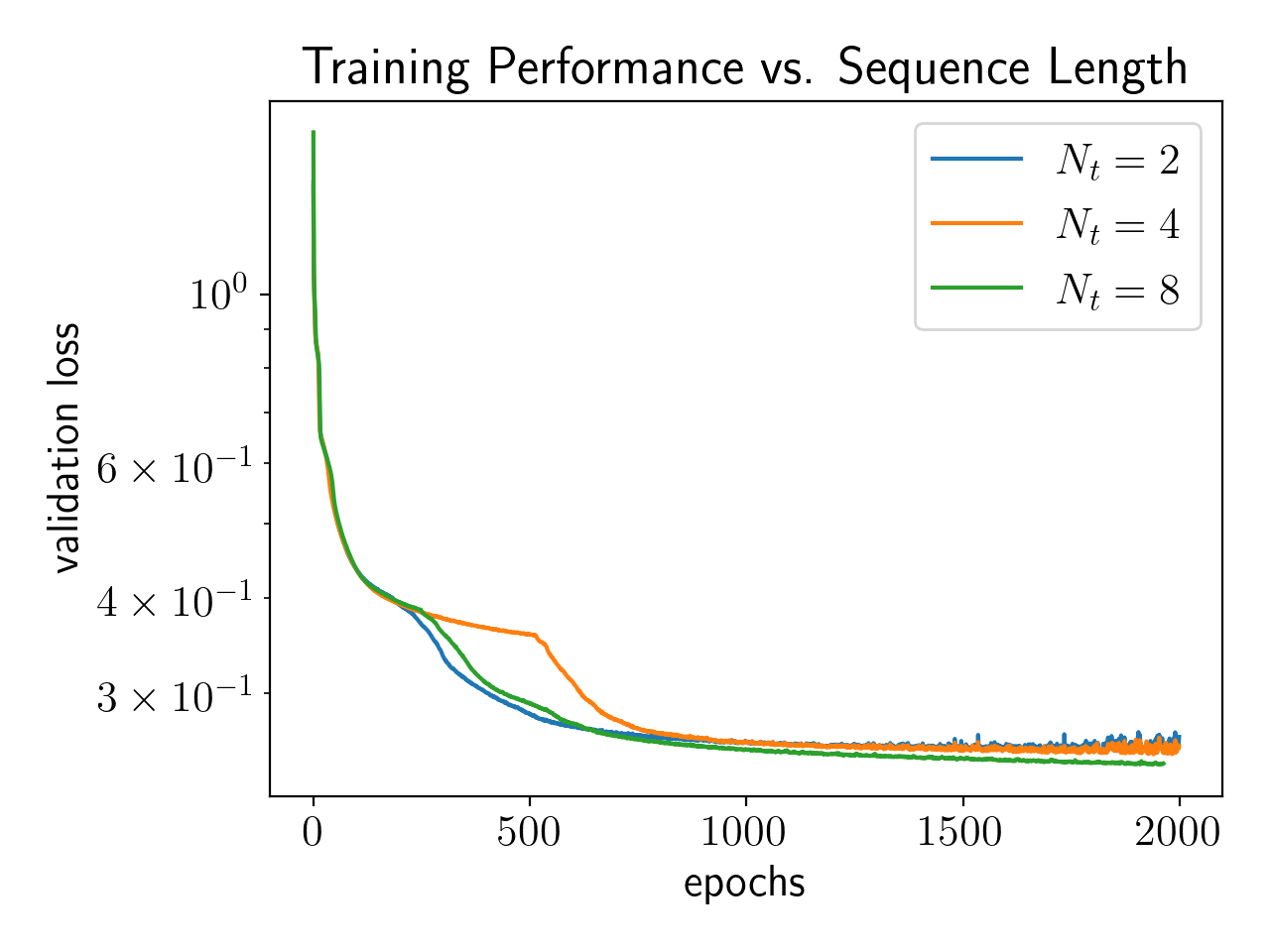}
        \caption{\label{fig:comp_b}}
    \end{subfigure}
    \caption{History of validation loss evaluated on the
    validation set consisting of 2,879 randomly chosen data samples
    for different training set sizes (left)
    and different sequence lengths (right).}
    \label{fig:data_comparisons}
\end{figure}
}

\begin{figure}[!htb]
    \centering

\resizebox{\textwidth}{!}{%
  \begin{tikzpicture}
  \useasboundingbox (0,.6) rectangle (15,5.75);  

  \draw(0, .5) node[anchor=south west] {\trimw{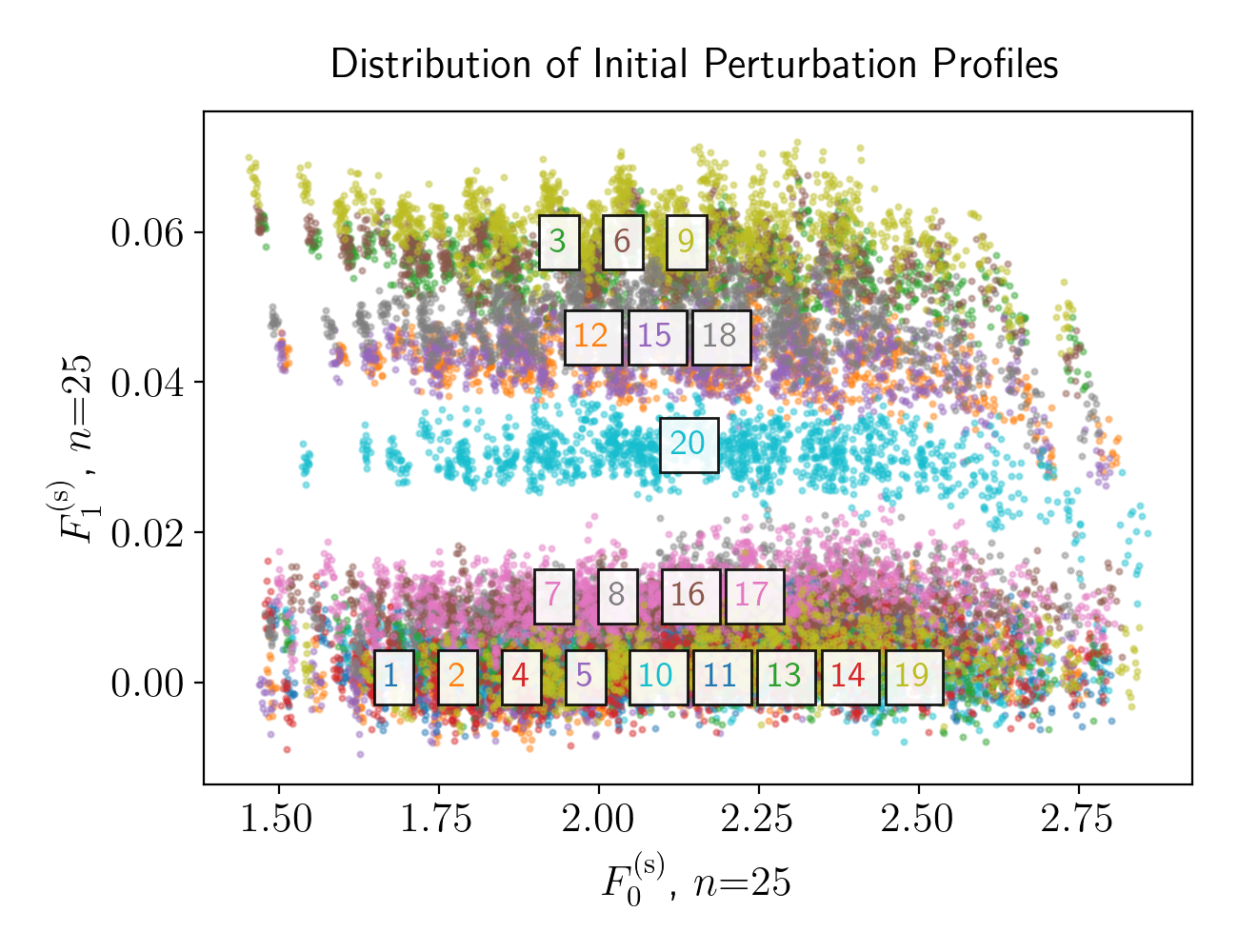}{7cm}{.0}{.0}{.0}{0}};

\draw(14.5, 2.5) node[anchor=south east, align=center, text width=7cm] {
\resizebox{1\columnwidth}{!}{%
    \begin{tabular}{|c|ccccc|}
    \hline
{\small Omitted Profile} & $v_{\rm impl}$ & $\Gamma_0$ & $s_1$ & $c_s$ & $c_v$ \\
\hline
{\small Profile 9} & 0.920 & 0.025 & 0.956 & 0.936 & -0.013 \\
{\small Profile 10} & 0.901 & 0.005 & 0.966 & 0.901 & 0.024 \\
{\small Profile 19} & 0.854 & 0.030 & 0.954 & 0.865 & 0.017 \\
{\small Profile 20} & 0.571 & 0.009 & 0.891 & 0.660 & 0.016 \\
\hline
    \end{tabular}}

};

    \draw(4.25, .6+0*4.75) node[] {(a)};
    \draw(11, 2+0*4.75) node[] {(b)};
    \draw(11, 4.75) node[] {Out of Sample Study};

  \end{tikzpicture}
}

    \bogus{
    \large{Out of Sample Study}

    \smallskip
        
    \resizebox{.4\columnwidth}{!}{%
    \begin{tabular}{|c|ccccc|}
    \hline
{\small Omitted Profile} & $v_{\rm impl}$ & $\Gamma_0$ & $s_1$ & $c_s$ & $c_v$ \\
\hline
{\small Profile 9} & 0.920 & 0.025 & 0.956 & 0.936 & -0.013 \\
{\small Profile 10} & 0.901 & 0.005 & 0.966 & 0.901 & 0.024 \\
{\small Profile 19} & 0.854 & 0.030 & 0.954 & 0.865 & 0.017 \\
{\small Profile 20} & 0.571 & 0.009 & 0.891 & 0.660 & 0.016 \\
\hline
    \end{tabular}}
}
    \caption{
    (a): Distribution of initial perturbation profiles in the 
    dataset as a function of the $0^{\rm th}$ and $1^{\rm st}$
    shock harmonics at $n=25$.
    (b): Correlation coefficients of initial velocity and 
    EOS parameter predictions 
    evaluated on profiles that
    were omitted from the training set.
    }
    \label{tab:omit}
\end{figure}

Next we consider a series of models
trained on all but one profile
and tested on the leftover profile.
The results of this study are summarized in 
Figure~\ref{tab:omit}b.
Despite being omitted from training, 
parameter estimation of 
profiles 9, 10, and 19 can be
performed with reasonable skill. 
However, profile 20 suffers from inaccuracies.
As can be seen by Figure~\ref{tab:omit}a,
in examining the zeroth and first shock
harmonics
of all the profiles as a point cloud in high dimensional space,
certain clusters emerge. 
Correspondingly, we observe that profiles 9, 10, 19 are clustered closely to other data while profile 20 forms it's own cluster, offering an explanation for the 
degraded performance.
The clusters are highly dependent the coefficients chosen in Table~\ref{tab:initialcoeffs}.

We now consider the sensitivity of the parameter estimation
with respect to the time frames considered for the features.
In this study, three
separate networks are trained using different
time frames of the input features, either
$\{25, 26, 27, 28\}$,
$\{30, 31, 32, 33\}$,
or
$\{35, 36, 37, 38\}$, 
and are compared to the baseline network
using the frames $\{25, 30, 35, 40\}$.
As seen in Table~\ref{tab:frames},
accuracy of the parameter estimates generally degrades 
when pushing the observable features to a later time.
Over time, the ratios between the 0$^{\rm th}$ harmonic and 
higher order harmonics of the shock grow in magnitude, 
so it is likely the network has a tougher time distinguishing between 
features at later times.
Despite this, using features spanning a larger time interval proves to be
the most successful in parameter estimation.

\begin{table}[htb]
    \centering
    \large{Information Content Study}

    \smallskip
        
    \resizebox{.8\columnwidth}{!}{%
    \begin{tabular}{|c|ccccccccccc|}
    \hline
frames & $F_1$ & $F_2$ & $F_3$ & $F_4$ & $F_6$ & $F_8$ & $v_{\rm impl}$ & $\Gamma_0$ & $s_1$ & $c_s$ & $c_v$ \\
\hline
$\{25, 30, 35, 40\}$  & 0.967 & 0.995 & 0.990 & 0.991 & 0.988 & 0.984 & 0.995 & 0.037 & 0.993 & 0.987 & -0.015 \\
$\{25, 26, 27, 28\}$ & 0.948 & 0.973 & 0.980 & 0.990 & 0.965 & 0.977 & 0.981 & 0.079 & 0.967 & 0.981 & 0.037 \\
$\{30,31,32,33\}$ & 0.921 & 0.992 & 0.965 & 0.988 & 0.947 & 0.977 & 0.955 & 0.044 & 0.927 & 0.971 & -0.004 \\
$\{35,36,37,38\}$ & 0.890 & 0.992 & 0.941 & 0.987 & 0.939 & 0.978 & 0.921 & 0.044 & 0.854 & 0.938 & 0.055 \\
\hline
    \end{tabular}}

    \caption{
    Correlation coefficients of EOS parameter and initial condition parameter predictions using
    different sets of time frames.
    }
    \label{tab:frames}
\end{table}

\subsection{Combining Parameter Estimation with Hydrodynamic Simulation}

In this section we demonstrate the utility of combining
our parameter estimation model with a hydrodynamic solver.
Since our model provides a two-way mapping between shock and edge
features and parameters, the learned forward model is limited 
in that it cannot provide additional information on 
state variables (e.g. density) and other characteristics 
of the solution (e.g. RMI topology) that can be
obtained from a full hydrodynamics solve.
We explore the feasibility of recovering the density, shock and edge
features, and the RMI topology using estimated parameters in a 
hydrodynamics solver.
We also utilize this section to explore the scenario of 
model mismatch by relating shock and edge features arising from 
different EOS models (e.g. Tillotson and Sesame)
to parameters of a Mie-Gr\"unieson model.

\bogus{
Given late-time shock harmonics, the network can predict the EOS 
and initial condition parameters.
These parameters can be used as an input to a hydrodynamic solver to obtain 
auxiliary information about the experiment, such as the full density field 
and peak-to-trough evolution of the RMI.
}

\begin{figure*}[htb]
    \centering
    \includegraphics[width=1\linewidth]{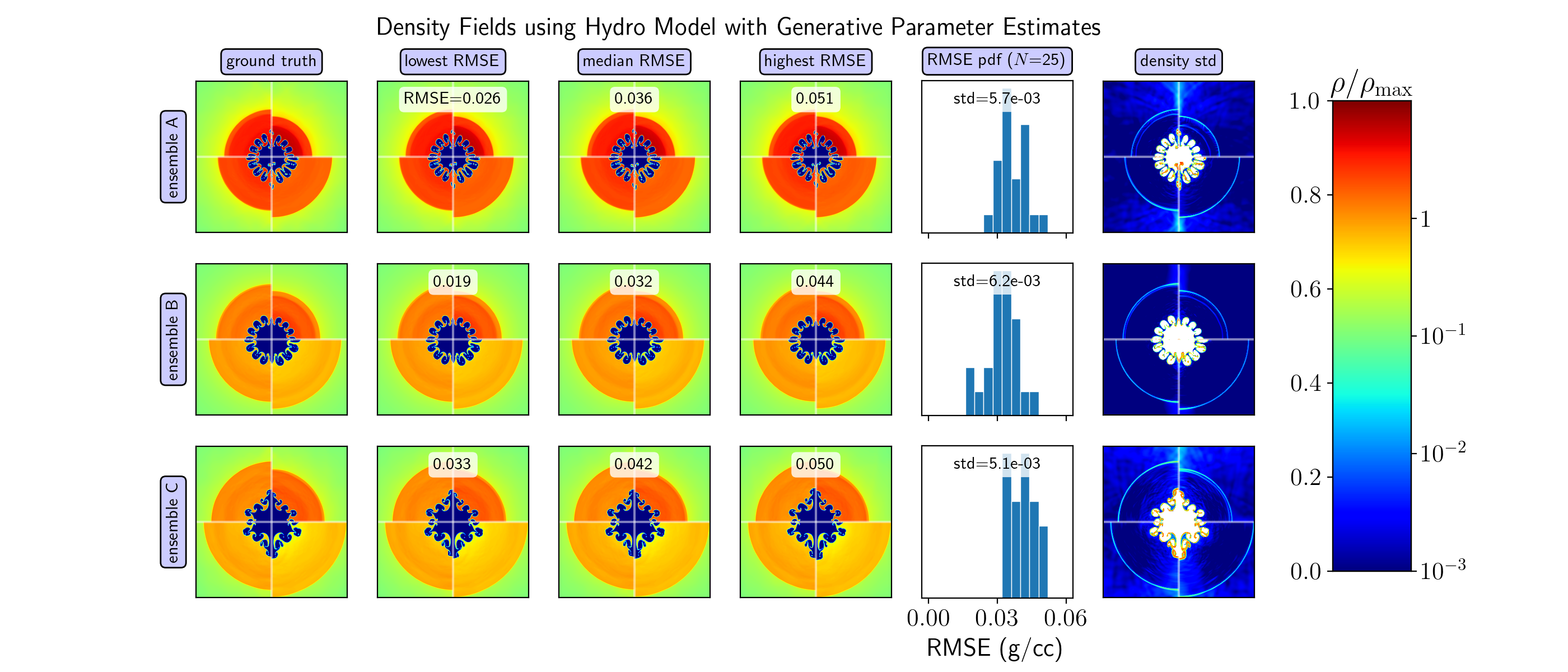}
    \caption{Comparison between three ground truth density 
    fields and ensembles of density fields obtained through 
    using estimated parameters in a hydrodynamics code.
    For each ensemble along the rows, this plot shows the ground truth density field, density fields corresponding to the lowest, median, and highest, RMSE, histogram of RMSEs, and standard deviation of density field.
    Each density field displays four quadrants corresponding to times $n=25, 30, 35, 40$.
    }
    \label{fig:compare_ensembles_density}
\end{figure*}

\begin{figure*}[htb]
    \centering
    \includegraphics[trim=5.5cm 5.5cm 5.5cm 5cm, clip, width=1\linewidth]{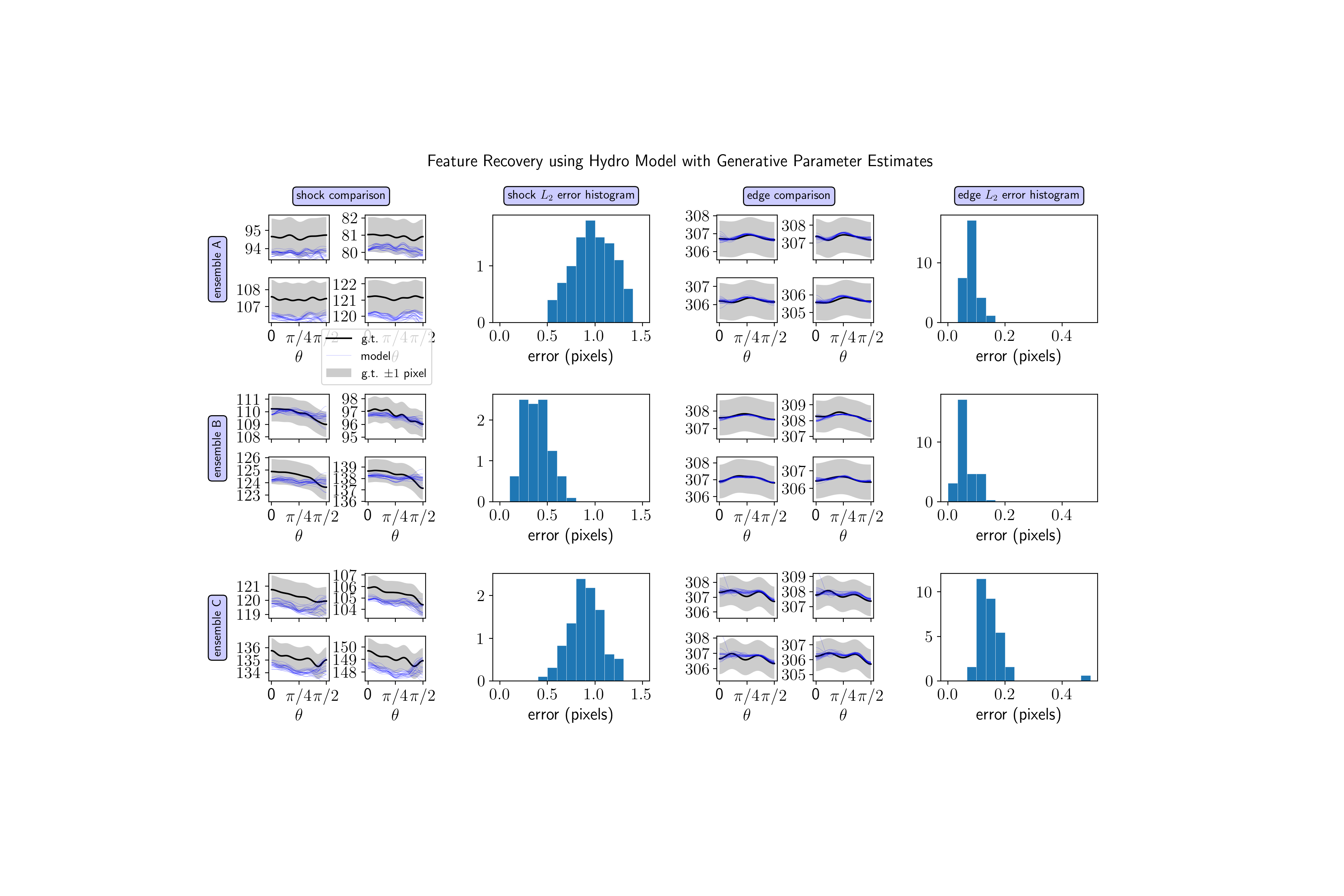}
    \caption{
    Comparison between three ground truth features (black lines) 
    and ensembles of features (blue lines) obtained through 
    using estimated parameters in a hydrodynamics code.
    For each ensemble along the rows, this plot shows 
    the comparison of the shock features, histogram of shock $L_2$ errors, comparison of edge features, and histogram of edge $L_2$ errors.
    Each feature comparison plot displays four quadrants corresponding to times $n=25, 30, 35, 40$.
    }
    \label{fig:compare_ensembles_feature}
\end{figure*}

\begin{figure*}[htb]
    \centering
    \includegraphics[width=.8\linewidth]{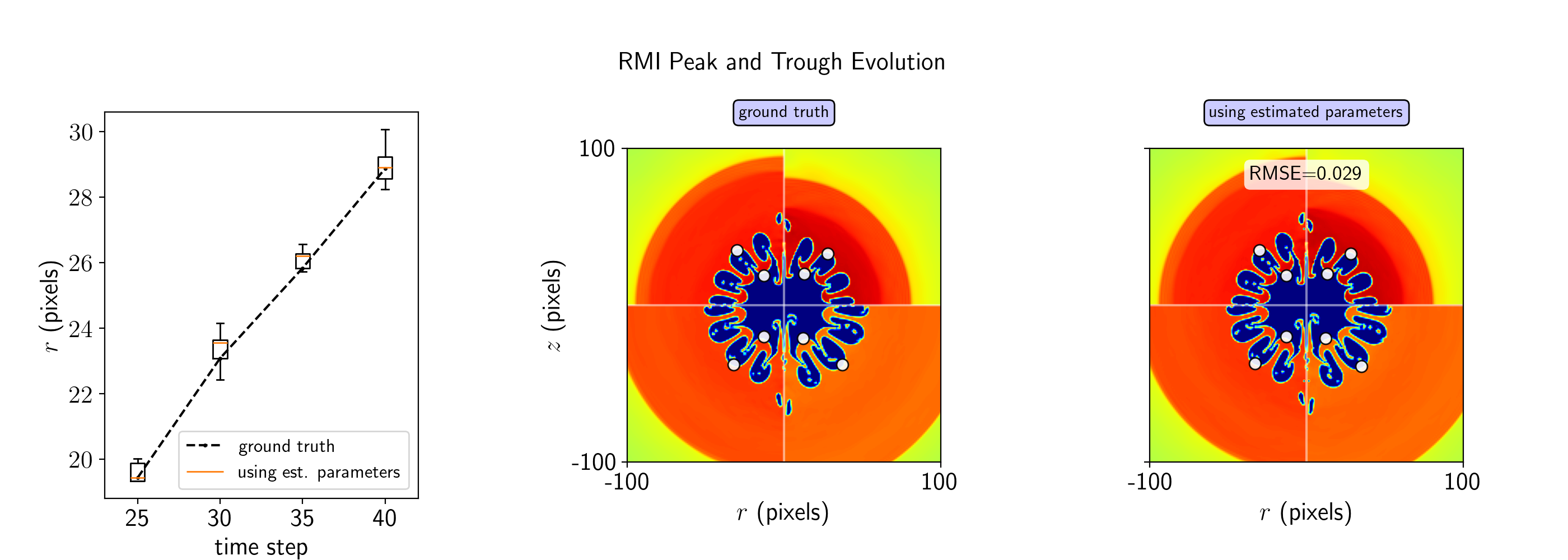}
    \caption{
    Example using estimated parameters 
    of ensemble A in Figure~\ref{fig:compare_ensembles_density}
    to predict peak-to-trough evolution of the RMI.
    Left:
  evolution of the maximum RMI peak-to-trough radial distance 
  for the ground truth and estimated parameters.
  Each density field displays four quadrants corresponding to time steps $25, 30, 35, 40$.
    Middle and right: density fields and identified
  peak and trough points of the RMI (white markers) 
  for the ground truth field
  and field obtained using estimated parameters.
  }
    \label{fig:ensemblebest}
\end{figure*}

In our first study, we consider three sets of features 
for which we generate ensembles ($N=25$) of predictions for each case
by sampling the generator latent space.
Each set of estimated parameters for each ensemble 
is then used as input into the hydrodynamic solver and the outputs
are compared to that of the ground truth.
Figure~\ref{fig:compare_ensembles_density} 
compares the ground truth density field
with the densities corresponding to the estimated 
parameters for the lowest, median, and highest
root mean squared error (RMSE) for each case
as well as the distribution of RMSE over each 
ensemble along with the standard deviation of the
density fields.
For each ensemble, the complex RMI surface 
is shown to be recovered with reasonable 
qualitative accuracy
and additionally the ensembles 
RMSEs are acceptably low and bounded.
The standard deviation of the density fields
have noticeable peaks near the shocks and RMI.
Figure~\ref{fig:compare_ensembles_feature}
compares the extracted shock and edge features for the same three 
ensembles.
The the $L_2$ errors for the edge feature are all captured accurately to within 0.5 pixels, the shock errors
are larger and bounded by 1.5 pixels for each ensemble.
Furthermore, Figure~\ref{fig:ensemblebest}
shows a qualitative and quantitative 
comparison between the peak and 
trough points of the RMI for ensemble A.
For this case, the peaks, troughs, 
and the growth of their distance
are all captured accurately.

\bogus{
\begin{wrapfigure}{r}{0.35\textwidth}
\centering
\vspace{-0.1in}
\includegraphics[width=0.23\textwidth]{figs/timing_weak_scaling.pdf}
\caption{Layer-parallel can accelerating training of neural ODEs on GPUs and mitigate the costs of training with increasing network depth (see~\cite{torchbraid-2024}).}
\label{fig:weak-scaling}
\end{wrapfigure}
}

\begin{figure*}[!htb]
    \centering
    \resizebox{\textwidth}{!}{%
  \begin{tikzpicture}
  \useasboundingbox (0,-0.5) rectangle (15,6.5);  

  \draw(7.5, 2.75) node[anchor=south] {\trimw{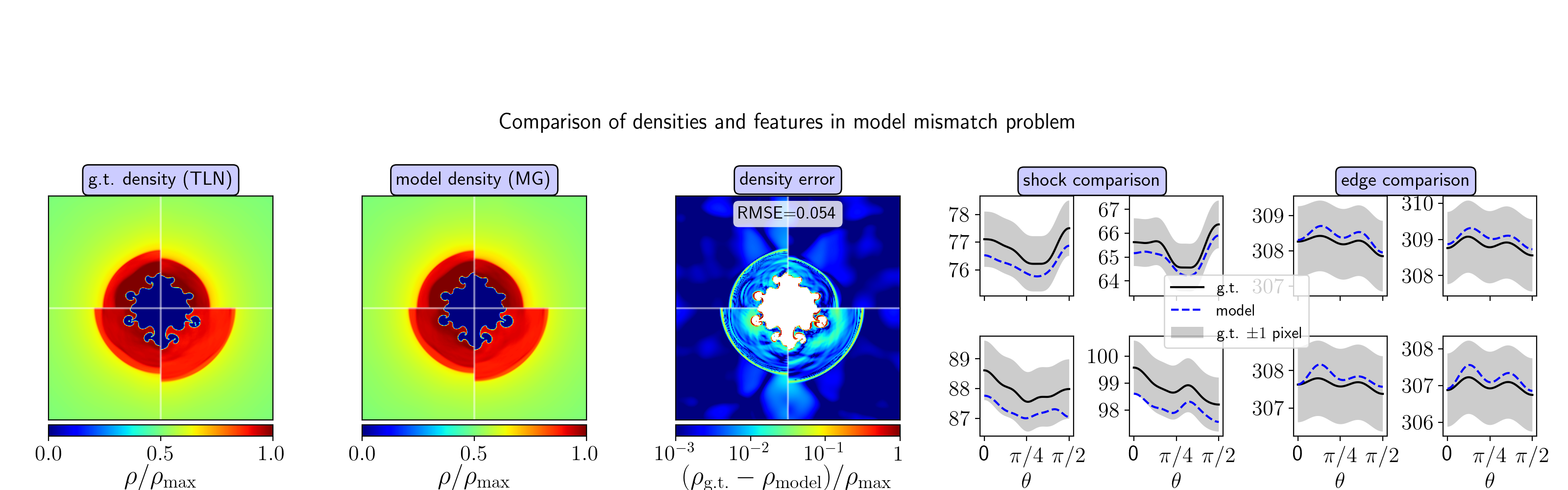}{14cm}{.0}{.0}{.0}{.3}};

  \draw(7.5, -1) node[anchor=south] {\trimw{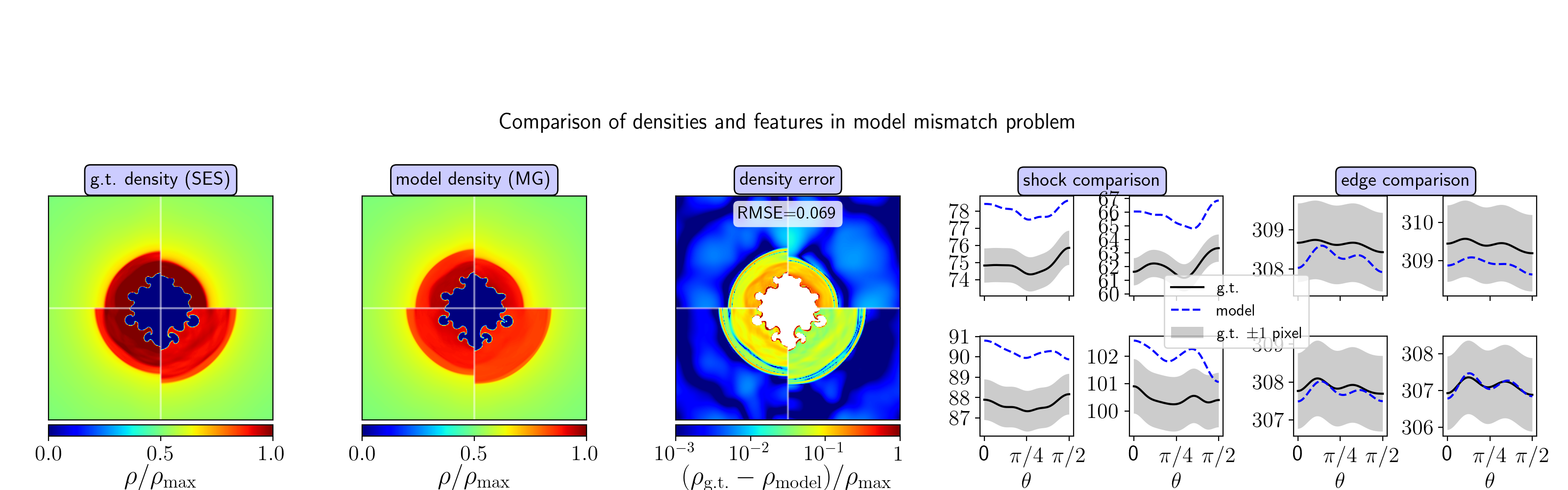}{14cm}{.0}{.0}{.0}{.3}};
    \draw(0.5, 4.5) node[] {(a)};
    \draw(0.5, 0.85) node[] {(b)};
    \draw(7.5, 6.2) node[] {Model Mismatch Study: Tillotson (TLN) EOS};
    \draw(7.5, 2.4) node[] {Model Mismatch Study: Sesame (SES) EOS};
  \end{tikzpicture}
}
\bogus{
     \includegraphics[width=1\linewidth]{paper_figs/eos_compare_tln_0.png}
    \includegraphics[width=1\linewidth]{paper_figs/eos_compare_ses_0.png}
    }
    \caption{
    Comparison between density fields and shock and edge features produced using the (a) Tillotson EOS and (b) Sesame EOS
    and the corresponding density fields and shock and edge features reconstructed using estimated parameters for the
    Mie-Gr\"uneison EOS model.
    Predicted shock and edge features are denoted by dotted blue lines and the ground truth features are denoted by black lines with a surrounding gray region depicting an error bar of $\pm$1 pixels.
    }
    \label{fig:tln_compare}
\end{figure*}

Next we explore the impact of model mismatch for our 
parameter estimation problem.
In our study, we chose to use Mie-Gr\"uneison (MG) as a guiding EOS model for our parameter estimation 
network.
However there will always be model mismatch and uncertainty
when comparing to experimental data or with simulations
utilizing alternative models.
In order to study the effect of model mismatch, 
we generate two sets of density time series 
using different EOS models -  
one using the Tillotson (TLN) EOS model~\cite{tillotson1962metallic}
and one using the Sesame (SES) EOS model~\cite{johnson1994sesame}.
For each series, we extract features corresponding to the
time frames $\{25, 30, 35, 40\}$
and input them into our network to predict 
corresponding MG parameters and initial conditions.
Figure~\ref{fig:tln_compare} shows comparisons 
between the ground truth density field 
and shock and edge features for both 
the TLN EOS and SES EOS
and the corresponding 
reconstructed density fields and and shock and edge features reconstructed using estimated parameters for the
MG EOS model.
For the TLN EOS, both the density field 
and shock and edge features are reconstructed
to a reasonable accuracy. 
However, while the material interface edge
is captured accurately for the SES EOS, 
large errors are present for the density field
and shock.
For both the TLN and SES EOS, 
the RMI topology predicted
using the MG EOS is 
qualitatively similar to the ground truth RMI
topology.
Errors in feature and density consistency 
due to model mismatch arise from two 
modalities.
First, the guiding EOS model on which parameter
estimates are based on must be sufficiently
expansive to approximate the unknown EOS model.
Second, the features produced by the unknown EOS
model must be adequately represented in the
training data used to optimize the ML model.
It is not surprising that our model
is able to able to perform better on 
simulations produced using the TLN EOS.
The MG and TLN EOS models are both globally 
described by a single equation, while the SES
is a tabular model that interpolates multiple
local models in different areas of the 
thermodynamic phase space.
Additionally, the features for the simulation 
using the SES model are outliers compared to the 
features used in training.

\bogus{
\begin{figure*}[ht]
    \centering
    \includegraphics[width=.48\linewidth]{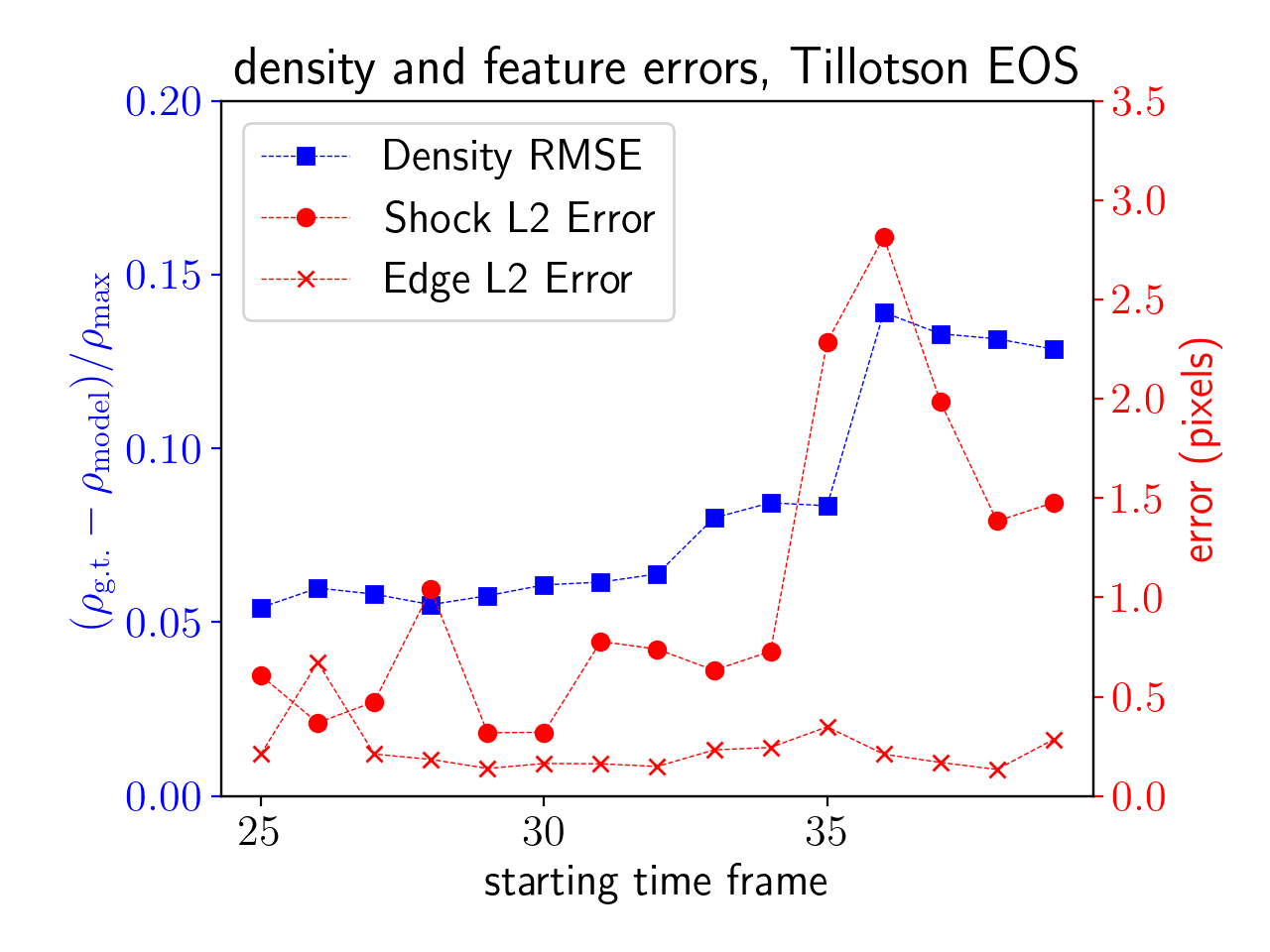}
    \includegraphics[width=.48\linewidth]{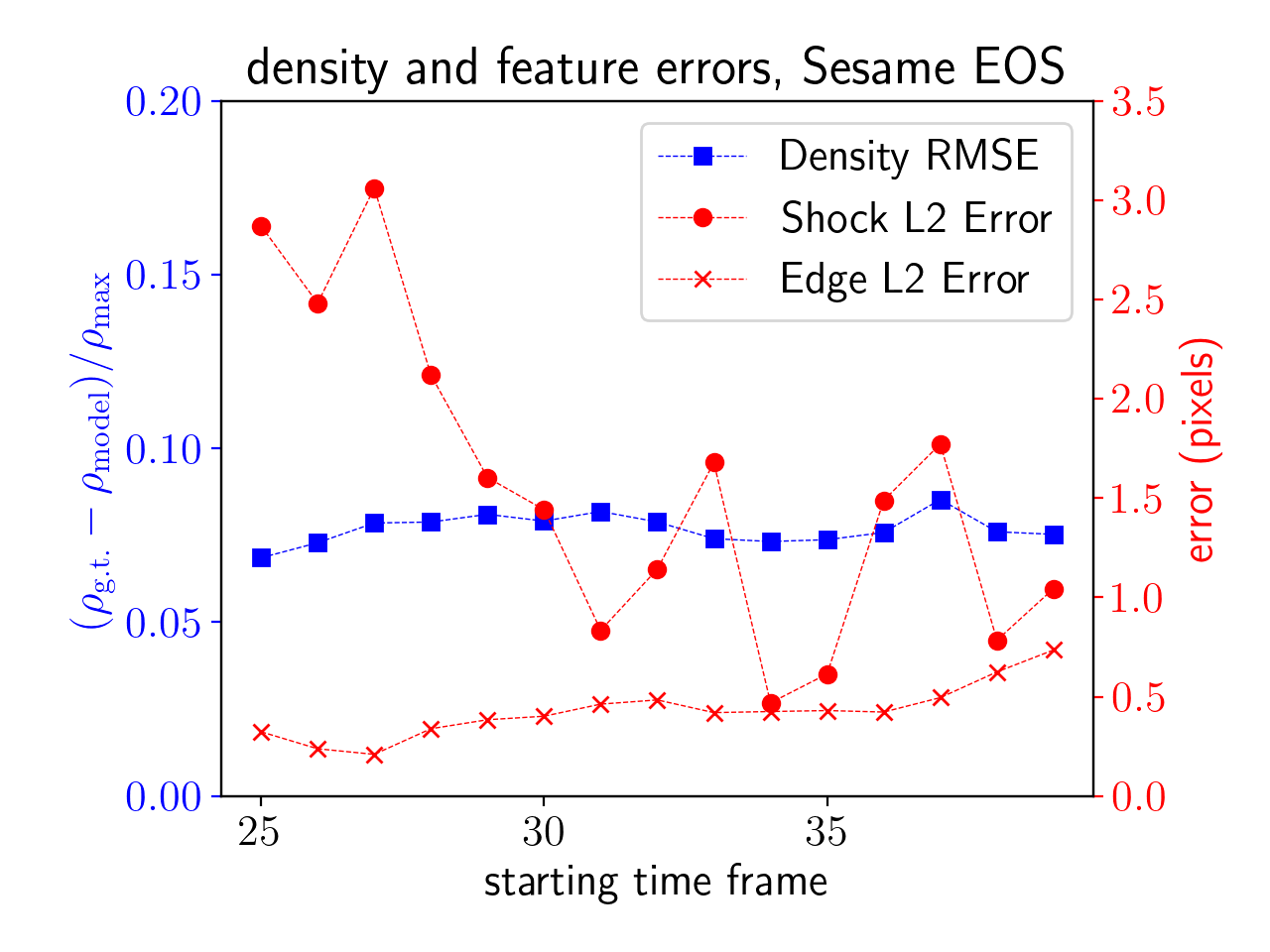}
    \caption{
    Errors between the density and features of the Tillotson and Sesame EOS-based simulations at different starting times and that of the simulation using estimated Mie-Gr\"uneisen-based parameters.
    }
    \label{fig:tln_compare_line}
\end{figure*}
}

\section{Discussion}

In this paper, we demonstrate a new 
machine learning (ML)-based approach
for recovering initial conditions and 
material parameters in ICF capsule implosions
that utilizes hydrodynamic features,
such as the outgoing shock profile and outer
material edge,
that are robustly identifiable in a noisy 
radiograph.
We propose that our method can be used
an experimental diagnostic to 
determine asymmetries in the drive 
that arise from the Ritchmeyer-Meshkov 
instability.
This experiment can be performed 
by using an inert gas in lieu of D-T inside
of the capsule or by inclusion of a suitable dopant to enable self-generated radiation to be imaged.
Our ML approach involves a pipeline
consisting of a radiographs-to-features
network (R2FNet) and 
a features-to-parameters network (F2PNet).
These networks can be trained independently 
and later combined during the testing stage.
F2PNet also contains a surrogate model for 
the forward hydrodynamic mapping between
parameters-to-features.

Our model problem consists of hydrodynamic simulations of 
an implosion of a nearly-spherical metallic shell. 
In our dataset, various equation of state (EOS) parameters 
and initial conditions on the surface
and initial velocity are varied to produce 
multiple realizations of density field time series.
For each simulation, we generated a 
time series of synthetic radiographs
and extracted shock and edge features in the 
form of cosine harmonic coefficients.
Through our results, we demonstrated that
our approach
is capable of recovering the EOS and
initial condition parameters with 
reasonable accuracy. 
We also show that the parameter estimates 
can be successfully used in a surrogate model 
for the forward problem to accurately 
estimate the shock and edge features.
Additionally, we show that the estimated 
parameters can be used in a (full-order) 
hydrodynamic solver to produce 
\emph{thermodynamically and hydrodynamically consistent}
density fields 
with acceptable quantitative accuracy in terms
of RMSE and errors in peak-to-trough distance 
of the RMI surface.
To the best of our knowledge, our framework is the first to 
recover \emph{thermodynamically and hydrodynamically consistent}
density reconstruction from noisy radiographs.

We also investigated the ability of the model
to estimate parameters when the 
reference simulation is 
generated with a different EOS model.
Our findings imply that a sufficiently 
expansive parameter model is capable of 
representing unknown models.
{
We note that there are other forms of model mismatch that 
manifest in experimental data, such as radiographic noise and scatter.
While the application of our approach to experimental data is 
the subject of our future work, we previously 
demonstrated in~\cite{hossain2022high}
that our method of using the shock and edge to reconstruct 
density fields was robust in the presence of 
out-of-distribution noise and scatter.
More recently, a comprehensive investigation was performed to
examine the impact of out-of-distribution noise and scatter~\cite{gautam2024learning}. The results again confirmed that the sparse features are
robust and consequently accurate density fields can be obtained from the dynamic sequence of these features
with significant out-of-distribution scatter and noise.
}

{
Since the F2PNet is generative, it is a useful tool in the
uncertainty analysis of predicted parameters.
Inspired by the conditional variational autoencoder (cVAE), 
the F2PNet architecture contains a decoder which 
applies a nonlinear transformation to features and a sample of 
multi-dimensional Gaussian noise to obtain a predicted distribution for
the parameters.
While the decoder is capable of producing multi-modal distributions, 
we largely observe that F2PNet trained on the dataset considered in this
paper predicts uni-modal distributions for the parameters. 
Typically, the characteristics of the predicted parameter distribution 
depend on the weighting of the KL divergence loss term and 
dimensionality of the noise in the latent space. 
However in our investigations, we observed 
insensitivity with respect to the choice of these meta
parameters and we therefore 
speculate that the observed uni-modal predictions are due to the data 
providing sufficient information content to closely identify the sensitive 
hydrodynamic parameters. 
Our speculation is also supported by 
our past work~\cite{Serino24}, where 
we investigated recovering
the density field in the vicinity of the Richtmeyer-Meshkov instability 
from the same hydrodynamic features dataset considered in this paper
using a generative cVAE-based vision transformer. 
In this work, we observed that sampling the latent space of the
generative network for a given feature 
produced a distribution of quantitatively close 
density fields with topologically similar Richtmeyer-Meshkov instabilities.
}

\appendix

\section{Cosine Coefficients for Inner Surface Perturbation Profile}

The coefficients of the cosine harmonic series of
the initial inner surface perturbation profile is scaled according to
$F_i = R_{\rm in} \bar{F}_i / 8,$ for
$i=0, \dots, 8$,
where $\bar{F}_0 = 8$, $\bar{F}_5=\bar{F}_7=0$,
and the rest of the coefficients are provided by Table~\ref{tab:initialcoeffs}.

\begin{table}
\begin{center}
  \resizebox{.97\columnwidth}{!}{%
\begin{tabular}{|c|c|c|c|c|c|c|c|c|}
\hline
profile& $\bar{F}_1$ & $\bar{F}_2$ & $\bar{F}_3$ & $\bar{F}_4$ & $\bar{F}_6$ & $\bar{F}_8$ \\
\hline
1 & 0 & 0 & 0 & 0 & 0 & 0.08\\
2 & 0 & 0 & 0 & 0.08 & 0 & 0\\
3 & 0 & 0.08 & 0 & 0 & 0 & 0\\
4 & 0 & 0 & 0 & 0 & 0 & 0.075\\
5 & 0 & 0 & 0 & 0.075 & 0 & 0\\
6 & 0 & 0.075 & 0 & 0 & 0 & 0\\
7 & 0 & 0.0075 & 0 & 0 & 0.0025 & 0.065\\
8 & 0.0075 & 0 & 0.0025 & 0.065 & 0 & 0\\
9 & 0.005 & 0.0657 & 0 & 0 & 0 & 0\\
10 & 0 & 0 & 0 & 0 & 0 & 0.06\\
\hline
\end{tabular}
\begin{tabular}{|c|c|c|c|c|c|c|c|c|}
\hline
profile& $\bar{F}_1$ & $\bar{F}_2$ & $\bar{F}_3$ & $\bar{F}_4$ & $\bar{F}_6$ & $\bar{F}_8$ \\
\hline
11 & 0 & 0 & 0 & 0.06 & 0 & 0\\
12 & 0 & 0.06 & 0 & 0 & 0 & 0\\
13 & 0 & 0 & 0 & 0 & 0 & 0.055\\
14 & 0 & 0 & 0 & 0.055 & 0 & 0\\
15 & 0 & 0.055 & 0 & 0 & 0 & 0\\
16 & 0 & 0.0075 & 0 & 0 & 0.0025 & 0.045\\
17 & 0.0075 & 0 & 0.0025 & 0.045 & 0 & 0\\
18 & 0.0051 & 0.0457 & 0 & 0 & 0 & 0\\
19 & 0 & 0 & 0 & 0.04 & 0 & 0\\
20 & 0 & 0.04 & 0 & 0 & 0 & 0\\
\hline
\end{tabular}
}
\end{center}
\caption{Scaled cosine coefficients for the initial shell profile used for
  each profile. 
  }
\label{tab:initialcoeffs}
\end{table}

\bogus{
\begin{table}
\begin{center}
\begin{tabular}{|c|c|c|c|c|c|c|c|c|}
\hline
profile& $\bar{F}_1$ & $\bar{F}_2$ & $\bar{F}_3$ & $\bar{F}_4$ & $\bar{F}_6$ & $\bar{F}_8$ \\
\hline
1 & 0 & 0 & 0 & 0 & 0 & 0.08\\
2 & 0 & 0 & 0 & 0.08 & 0 & 0\\
3 & 0 & 0.08 & 0 & 0 & 0 & 0\\
4 & 0 & 0 & 0 & 0 & 0 & 0.075\\
5 & 0 & 0 & 0 & 0.075 & 0 & 0\\
6 & 0 & 0.075 & 0 & 0 & 0 & 0\\
7 & 0 & 0.0075 & 0 & 0 & 0.0025 & 0.065\\
8 & 0.0075 & 0 & 0.0025 & 0.065 & 0 & 0\\
9 & 0.005 & 0.0657 & 0 & 0 & 0 & 0\\
10 & 0 & 0 & 0 & 0 & 0 & 0.06\\
11 & 0 & 0 & 0 & 0.06 & 0 & 0\\
12 & 0 & 0.06 & 0 & 0 & 0 & 0\\
13 & 0 & 0 & 0 & 0 & 0 & 0.055\\
14 & 0 & 0 & 0 & 0.055 & 0 & 0\\
15 & 0 & 0.055 & 0 & 0 & 0 & 0\\
16 & 0 & 0.0075 & 0 & 0 & 0.0025 & 0.045\\
17 & 0.0075 & 0 & 0.0025 & 0.045 & 0 & 0\\
18 & 0.0051 & 0.0457 & 0 & 0 & 0 & 0\\
19 & 0 & 0 & 0 & 0.04 & 0 & 0\\
20 & 0 & 0.04 & 0 & 0 & 0 & 0\\
\hline
\end{tabular}
\end{center}
\caption{Scaled cosine coefficients for the initial shell profile used for
  each profile. 
  }
\label{tab:initialcoeffs}
\end{table}
}

\section{Radiograph-to-Features Network (R2FNet)}
\label{sec:r2f}

\begin{figure}[tbh]
    \centering
    {\large Radiograph-to-Features Network (R2FNet)}
    \includegraphics[width=1\linewidth]{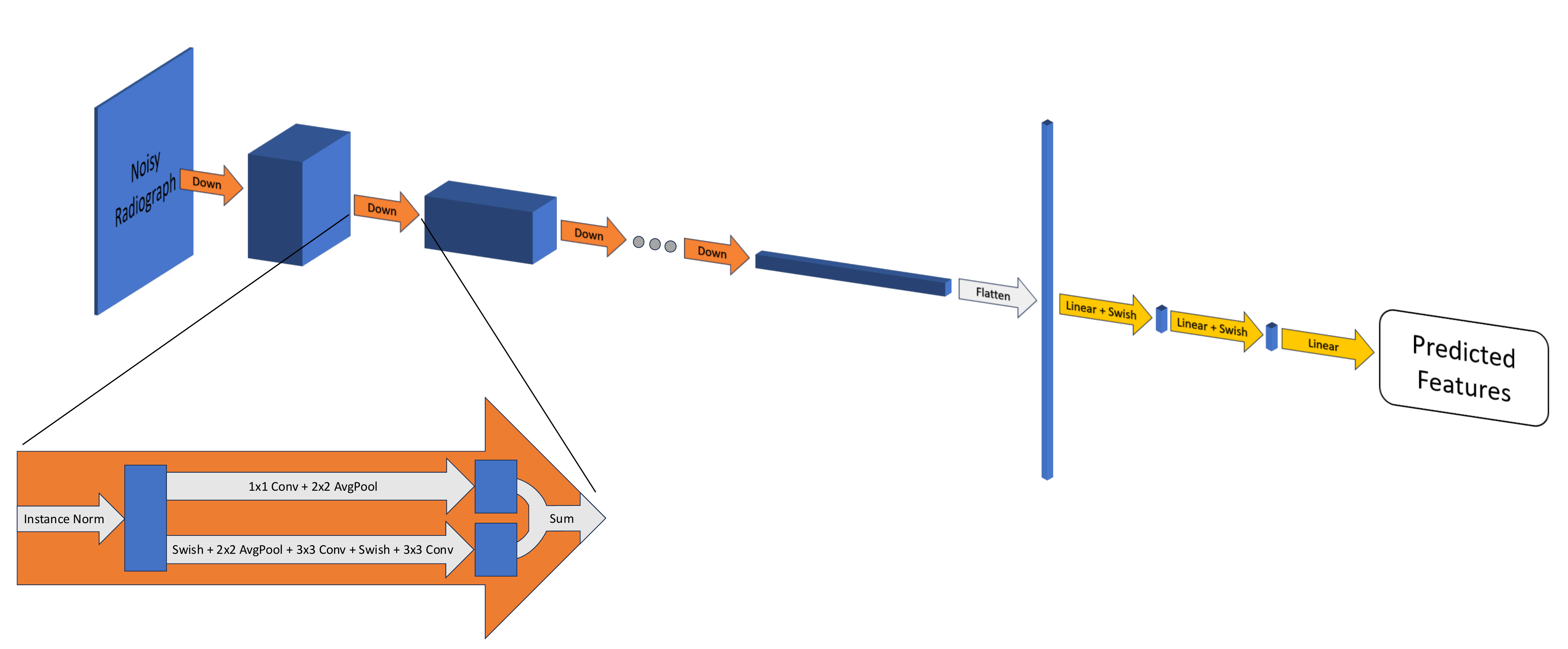}
    \caption{Diagram depicting the radiograph-to-features network (R2FNet). 
    The input is a sequence of noisy radiographs and the output is 
    a corresponding sequence of predicted shock and edge features.}
    \label{fig:r2fnet}
\end{figure}

The radiograph-to-features network (R2FNet) is illustrated in 
Figure~\ref{fig:r2fnet}
and is built entirely from 
convolutions and fully connected linear layers.
The first component of the architecture is a convolutional encoder which compresses the series of radiographs from their large ambient dimension to a much smaller latent dimension. 
The downsampling blocks used in this architecture are identical to the residual blocks used in~\cite{deblurring_via_sr}, and are very similar to those found in BigGAN~\cite{big_gan}. The network input consists of the four radiographs stacked together so that each timestep is represented by one input channel. The first downsampling block then increases the channel dimension to 64. Each further downsampling block adds an additional 64 channels, and all the downsampling blocks reduce the spatial dimensions by a factor of 2. In the downsampling blocks, the first 3${\times}$3 convolution increases the number of channels. Both networks use 7 downsampling blocks.
The outputs of the encoder are then flattened into a vector. 
Subsequently, there are 3 linear layers - 
the first of which reduces the dimension by a factor of 16, the second maintains this dimension, and the final layer outputs the parameter predictions.

Because the cosine harmonic coefficients of the features of interest have very different absolute scales, they are normalized before being used as targets for training the network. We apply z-score normalization to each coefficient by subtracting its population mean and dividing by its population standard deviation. The loss function used for training is a combination of the mean squared error of the predicted (normalized) coefficients and the mean squared error of the radius of the shock and radius of the edge, sampled on a grid of 500 equally spaced angles $\theta \in [0, \pi/2)$. The R2FNet is optimized using Adam~\cite{adam}, with a learning rate of $10^{-5}$, and a batch size of 8 using eight NVIDIA GeForce RTX 2080 Ti GPUs. The R2FNet achieved its minimum validation loss after approximately 8 hours.

\section{Features-to-Parameters Network (F2PNet)}
\label{sec:f2p}

\begin{figure*}[!htb]
    \centering
    \includegraphics[width=.9\linewidth]{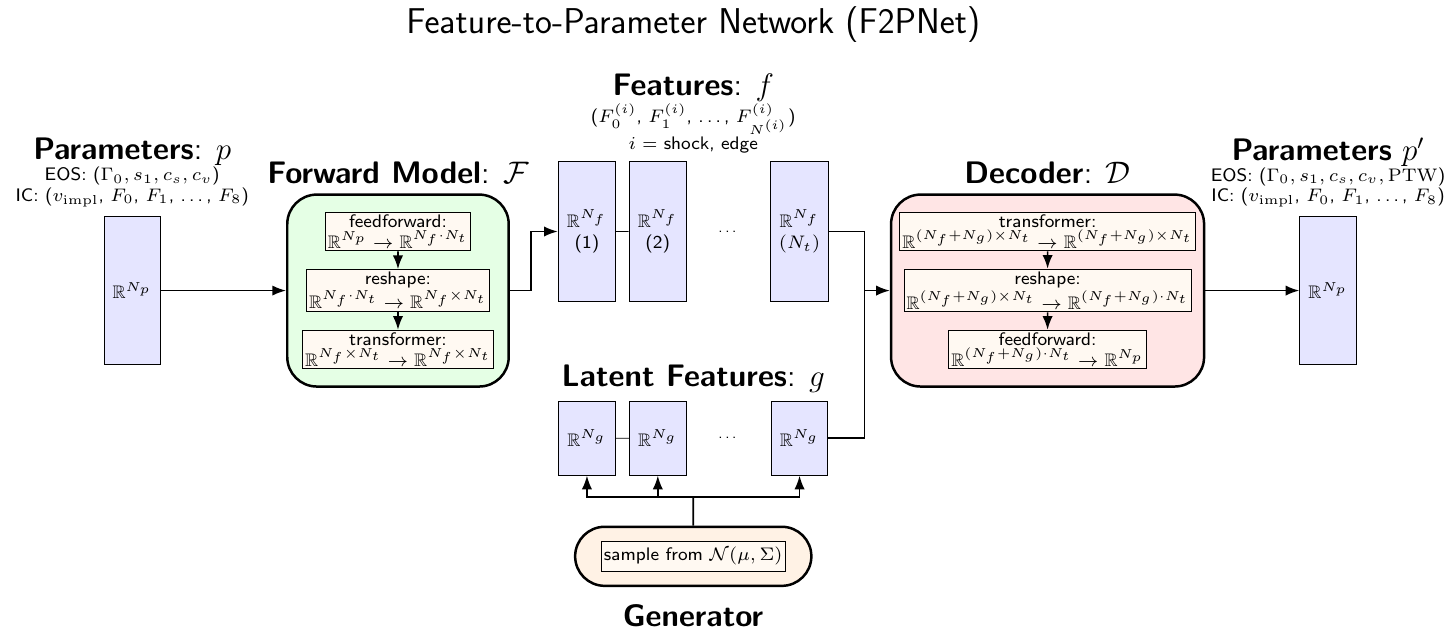}
    \caption{Diagram depicting the features-to-parameters network (F2PNet). 
    The EOS and IC parameters represent the input to the forward model and output of the decoder. 
    The late time shock and edge features are outputs of the encoder. The late time shock and edge features and the latent features randomly sampled from the generator are used as inputs to the decoder.}
    \label{fig:tcvae}
\end{figure*}

Our features-to-parameters network (F2PNet)
combines ideas from the conditional variational autoencoder (cVAE)~\cite{kingma2013auto,cinelli2021variational,kingma2014semi} 
and the transformer~\cite{Vaswani17},
for estimating initial condition (IC) and EOS parameters based on 
observed outgoing shock and outer edge features.
The F2PNet architecture is summarized by 
Figure~\ref{fig:tcvae}.
F2PNet consists of a forward model,
which is a surrogate model for 
the hydrodynamics operator mapping parameters
to features, 
and a generative parameter estimator, 
which represents the decoder of a cVAE. 
For all network inputs and outputs, 
each EOS parameter and IC parameter are linearly scaled to 
lie in the range $[0, 1]$ 
and each shock feature is scaled according to 
its corresponding mean and standard deviation 
in the training set.

Consider parameters, $p\in\mathbb{R}^{N_p}$,
late-time shock and edge features at $N_t$ times, $f\in\mathbb{R}^{N_f\times N_t}$,
and generator features, $g\in\mathbb{R}^{N_g \times N_t}$.
The forward model, $\mathcal{F}$, represents a surrogate forward model, 
predicting late-time shock and edge features from parameters.
The first layer is a fully-connected feedforward neural network
which outputs data that is reshaped into $\mathbb{R}^{N_f \times N_t}$.
A transformer network, based on \cite{Vaswani17}, is then applied to 
obtain late-time shock and edge features.
The generator draws vectors $g\in \mathbb{R}^{N_f\cdot N_t}$ 
from the probability
distribution $\mathcal{N}(\mu, \Sigma)$,
where $\mu \in \mathbb{R}^{N_f\cdot N_t}$ and $\Sigma\in \mathbb{R}^{(N_g\cdot N_t) \times (Ng\cdot Nt)}$ are learned parameters.
The decoder, $\mathcal{D}$, performs the parameter estimation task given the 
late-time shock and edge features and randomly generated features.
The first layer is a transformer network which outputs
data in a shape $\mathbb{R}^{N_t \times (N_f + N_g)}$.
After reshaping, a fully-connected feedforward neural network is then applied to obtain
the parameters.

The model architecture weights are
trained using a loss of the form
$
\mathcal{L} =
\mathcal{L}_{\rm decoder}
+ \mathcal{L}_{\rm forward}
+ \mathcal{L}_{\rm consistency}
+ \mathcal{L}_{\rm KL},
$
%
where 
\begin{align*}
\mathcal{L}_{\rm decoder} = \frac{1}{N_d}\sum_{i=1}^{N_d} \|\mathcal{D}(f_i; g_i) - p_i\|^2, 
\quad
\mathcal{L}_{\rm forward} = \frac{1}{N_d}\sum_{i=1}^{N_d} 
L_2(\mathcal{F}(p_i), f_i)^2,
\quad
\mathcal{L}_{\rm consistency} = \frac{\alpha}
{N_d}\sum_{i=1}^{N_d} 
L_2(\mathcal{F}(\mathcal{D}(f_i; g_i)), f_i),
\end{align*}
and 
$\mathcal{L}_{\rm KL} =
D_{\rm KL}\left(\mathcal{N}(\mu, \Sigma) \; \| \; \mathcal{N}(0, I)\right)$.
$\mathcal{L}_{\rm decoder}$
represents the mean squared error between the
decoder's prediction of parameters 
and the ground truth,
%
$\mathcal{L}_{\rm forward}$
represents the squared $L_2$ error between the shock and edge 
curves defined by the forward model's prediction of 
feature coefficients and that of the 
ground truth,
%
$\mathcal{L}_{\rm consistency}$
represents a self-consistency loss using
the same error metric as
$\mathcal{L}_{\rm forward}$, but instead using the 
shock and edge curves reconstructed by applying the 
forward model to the decoder's parameter estimates of the ground
truth features, and
$\mathcal{L}_{\rm KL}$ is the 
KL divergence between the generator probability distribution 
and a standard Gaussian distribution.
The combination of $\mathcal{L}_{\rm decoder}$
and $\mathcal{L}_{\rm KL}$ represents the 
traditional VAE loss~\cite{kingma2013auto} while 
the addition of 
$\mathcal{L}_{\rm forward}$ and
$\mathcal{L}_{\rm consistency}$
encourages 
accuracy and consistency of the forward model.
In our study,
we first pre-train the architecture without the 
self-consistency term ($\alpha = 0$)
to obtain a reasonable initial set of weights for the 
encoder and decoder,
then later perform a second round of training
using all of the terms ($\alpha = 1$).

In summary, F2PNet combines the
strengths of the transformer and
conditional variational autoencoder to estimate parameters
from temporal sequences of shock and edge features.
The transformer provides the
ability to capture long range temporal dependencies.  
Results are presented for an architecture that uses
two transformer blocks in the encoder and decoder,
each with 8 heads, $H=8$,
a latent dimension of $k=64$,
and a feedforward neural network with inner dimension 2048
and $\tanh$ activation functions.
The feedforward neural networks in the encoder and decoder 
each have two layers with a hidden dimension of 200.

In our study, we also investigated the use of different network architectures,
including networks without attention mechanisms.
We discovered that applying attention greatly improves the
prediction accuracy of the parameters and therefore 
we chose to make this a key component of F2PNet.
Additionally, we investigated a probabilistic neural network, 
which models the posterior as a multivariate Gaussian 
distribution.
However, since the posterior is generally a complex probability distribution,
we decided to pursue a VAE-based architecture, which is capable of 
representing non-Gaussian distributions.

\section*{Data availability}
The datasets analyzed during the current study are available from the corresponding author on reasonable request.

\bibliography{refs.bib}{}

\section*{Acknowledgements}

This work was supported by the Laboratory Directed Research and Development program of Los Alamos National Laboratory under project number 20230068DR.

\section*{Author contributions statement}

D.A.S, E.B., M.K., B.S.S., and B.N. conceived the parameter estimation approach.
O.K. developed the feature extraction algorithm.
D.A.S., E.B., and B.N. conducted the approach.
D.A.S., E.B., M.K., and B.S.S. analyzed the results.
All authors reviewed the manuscript.

\section*{Competing interests}

The authors declare no competing interests.

\end{document}